\documentclass{article}

\usepackage{moreverb}

\usepackage[colorlinks,bookmarksopen,bookmarksnumbered,citecolor=red,urlcolor=red]{hyperref}
\usepackage[a4paper,body={17cm,24cm},top=3.cm]{geometry}

\usepackage[utf8]{inputenc}
\usepackage[T1]{fontenc}
\usepackage[english]{babel}
\usepackage{amsmath,amsfonts,amssymb,amsthm}
\usepackage{graphicx,color,multirow}
\theoremstyle{remark}

\newcommand{\A}{\ensuremath{\mathcal{A}}}
\newcommand{\R}{\ensuremath{\mathbb{R}}}
\newcommand{\C}{\ensuremath{\mathbb{C}}}
\newcommand{\Real}{\ensuremath{\operatorname{Re}}}
\newcommand{\Imag}{\ensuremath{\operatorname{Im}}}

\newcommand{\rmi}{\,\mathrm{i}}
\newcommand{\dV}[0]{\,\mathrm{d}V}

\newcommand{\dth}[0]{\,\mathrm{d}\theta}

\newcommand{\jump}[1]{[\mspace{-2 mu} [ #1 ]\mspace{-2 mu} ]}
\newcommand{\mean}[1]{\{\mspace{-6 mu} \{ #1 \}\mspace{-6 mu}\}}

\newcommand{\At}{\tilde{\mathcal{A}}}
\newcommand{\tV}{\tilde{\mathbf{V}}}

\newcommand{\sH}{\mathcal{V}}
\newcommand{\sS}{\mathcal{S}}
\newcommand{\parti}{\mathcal{T}}
\newcommand{\setF}{\mathcal{F}}
\newcommand{\sST}{\sS^\parti}

\newcommand{\massC}{\mathbf{M}_\mathcal{C}}
\newcommand{\massh}{\mathbf{M}_h}
\newcommand{\proj}{\mathbf{X}}

\newcommand\BibTeX{{\rmfamily B\kern-.05em \textsc{i\kern-.025em b}\kern-.08em
T\kern-.1667em\lower.7ex\hbox{E}\kern-.125emX}}

\newcommand{\basA}{\mathbf{A}}
\newcommand{\E}{^{E}}
\newcommand{\ET}{^{E^H}}
\newcommand{\basPA}{\mathbf{P}_\mathbf{A}}
\newcommand{\basPAh}{\mathbf{P}_{\mathbf{A},h}}
\newcommand{\bal}{\boldsymbol{\alpha}}

\begin{document}

\author{L. Kovalevsky $^{1}$, P. Gosselet$^2$ \\
  (1) Department of Engineering, University of Cambridge, \\Trumpington Street, Cambridge CB2 1PZ, UK \\
(2) LMT-Cachan, ENS-Cachan/CNRS/Pres UniverSud Paris, \\61 avenue du Président Wilson, 94235 Cachan, France}

\title{A quasi-optimal coarse problem and an augmented Krylov solver for the Variational Theory of Complex Rays}

\maketitle
\begin{abstract}
{
The Variational Theory of Complex Rays (VTCR) is an indirect Trefftz method designed to study systems governed by Helmholtz-like equations. It uses wave functions to represent the solution inside elements, which reduces the dispersion error compared to classical polynomial approaches but the resulting system is prone to be ill conditioned. This paper gives a simple and original presentation of the VTCR using the discontinuous Galerkin framework and it traces back the ill-conditioning to the  accumulation of eigenvalues near zero for the formulation written in terms of wave amplitude. The core of this paper presents an efficient solving strategy that overcomes this issue. The key element is the construction of a search subspace where the condition number is controlled at the cost of a limited decrease of attainable precision.  An augmented LSQR solver is then proposed to solve efficiently and accurately the complete system. The approach is successfully applied to different examples.}
\end{abstract}
\medskip 

\noindent Final paper is accepted in International Journal For Numerical Methods in Engineering (doi: 10.1002/nme.5190)

\section{Introduction}

In the last decades, the use of numerical simulation techniques in design, analysis and optimization of systems has become an indispensable part of the industrial design process. 
The most used computer aided engineering tool is the standard Galerkin Finite Element Method (FEM~\cite{zienkiewicz1977}). It is based on the use  of continuous, piecewise-polynomial shape functions supported by a mesh. It applies particularly well on the Poisson equation where a coercive formulation naturally arises. 

In the case of the Helmholtz equation, the straightforward formulation is not positive, and then a fine discretization is required in order to limit the dispersion and pollution errors induced by the non-verification of the governing partial differential equations \cite{Deraemaeker1999}. Robust FEM approaches imply to use adapted formulation and elements \cite{Moiola2014}.

Alternative techniques exist based on the discontinuous Galerkin methods (DG)~\cite{dg12} which allow the discontinuity of the shape functions between elements, so that any type of shape function can be used. In particular, the Trefftz-DG methods use basis functions that are locally (i.e. inside each mesh element) solutions of the relevant governing partial differential equations (PDEs). In most cases, coercivity can be ensured for these methods, which moreover lead to  smaller dispersion error than the finite element method (see  \cite{gittelson2014dispersion} for a discussion on the error in Trefftz-DG approaches).
Such approaches include, for example, the ultra-weak variational formulation (UWVF)~\cite{cessenatu}, the discontinuous enrichment method~\cite{farhatu}, the wave-based method~\cite{desmetu} and the variational theory of complex rays (VTCR)~ \cite{ladevezeu}. The main differences among the various Trefftz methods lie in the treatment of the boundary conditions and of the continuity conditions between elements, in the type of waves used in the admissible space and on the chosen discrete unknowns. For a given level of accuracy, all these methods lead to a substantially smaller algebraic system than one would obtain using the standard FEM. However, these methods often suffer from an ill-conditioned algebraical system (see for example~\cite{desmetu,strouboulist,riou2008,kovalevsky12}) even after scaling the different terms of the formulation \cite{huttunen2008interaction,farhatu}. 
In order to control the condition number it was proposed in \cite{huttunen2002computational} to iteratively enrich the basis of wave functions until the condition number of the element matrix becomes too poor. A close idea was proposed in \cite{kovalevsky12} for the VTCR. 
\smallskip

This paper is dedicated to the study of the VTCR for acoustic problems, which is an indirect Treffz-DG method where Herglotz wave functions are used to represent the solution inside elements and where inter-elements conditions are dealt with by an anti-hermitian formulation. In Section~\ref{VTCR} we give an original and pedagogic presentation of the VTCR and we prove that under general assumptions the formulation in terms of pressure is coercive. In Section~\ref{OMG} we show that the problem being set in terms of amplitudes via the (compact) Herglotz operator leads to non coercivity (even if still sign definite) because eigenvalues accumulate near zero. This phenomenon causes the bad conditioning of the discrete system. We then propose to build a search space where coercivity is restored at the price of a small decrease of the precision. In Section~\ref{sec:practical}, the construction of the subspace is detailed, and it is proposed to use it as the coarse grid for an augmented Krylov solver. 

{The subspace that we build is quasi-optimal in the sense that, starting from an initial search space, it is a controlled approximation of the largest subspace where a chosen level of coercivity (or condition number) is preserved.}

Note that this strategy is different from previous work as \cite{huttunen2002computational}, it does not set up an upper limit for the number of waves, but selects all combinations of waves which result in pressure fields containing at least a certain amount of energy. This collection of waves spans what we call in the following the optimized subspace. The threshold used to distinguish sufficiently energetic modes from others is a parameter of the method which can be replaced by a criterion on the proportion of the total energy which should be present in the subsystem. Note that the final precision is not limited thanks to the use of the augmented LSQR solver, in that context the subspace is referred to as the coarse subspace.  Two assessments are given, the first one (Section~\ref{subsection: scattering of a cylinder}) possesses an analytical solution which enables us to fully understand the properties of the selected subspace and of the augmented solver; the second one (Section~\ref{subsection: car}) is more realistic and enables us to illustrate the potential of the method in terms of computational performance.

\section{The Variational Theory of Complex Rays  for Helmholtz problems}
\label{VTCR}

\subsection{Reference problem}

Let us consider a bounded acoustic cavity $\Omega\subset\R^d$ filled with a fluid characterized by its speed of sound $c$, its density $\rho$ and its damping coefficient $\eta$. A source term $f$ is given in $\Omega$. The boundary $\partial\Omega$ is partitioned into three parts: $\partial_p\Omega$ where the pressure $\tilde{p}_d$ is prescribed, $\partial_v\Omega$ where the velocity $\tilde{v}_d$ is prescribed and $\partial_Z\Omega$ where a Robin condition $\tilde{h}_d$ is imposed, the impedance being written $Z$ ($Z\neq 0$ on $\partial_Z\Omega$). Assuming all excitations are time-harmonic with a given circular frequency $\omega$, the complex acoustic pressure $p$ in $\Omega$ solves the following boundary-value problem for the Helmholtz equation:
{
\begin{equation}\label{equation: reference problem}
  \textrm{Find $p$ such that} \qquad
\left |
\begin{aligned}
  \Delta p + k^2 p &= f \quad &&\textrm{in $\Omega$} && \text{(a)}\\
  Z p +   \frac{\partial p}{\partial n} &= \tilde{h}_d \quad &&\textrm{over $ \partial_Z \Omega$} && \text{(b)} \\
  p  &= \tilde{p}_d \quad &&\textrm{over $ \partial_p \Omega$} && \text{(c)} \\
 \frac{\partial p}{\partial n}  &= \tilde{v}_d \quad &&\textrm{over $ \partial_v \Omega$} && \text{(d)} \\
\end{aligned} \right.
\end{equation}
}
where $k = (1 + \rmi \eta)(\omega/c)$ is the wave number, $n$  is the outward normal to $\partial \Omega$, $\rmi = \sqrt{-1}$ is the imaginary unit. For physical consideration, it is assumed that $\eta\geqslant 0$ and $\Real (Z)\geqslant 0$.

To apply the VTCR, an homogeneous equation is required. The reference system is then modified by the introduction of a solution
$p^c$ to~\eqref{equation: reference problem}(a) under the form $p^c(\mathbf{x})=\int_{\Omega} \rho(\mathbf{y}) f(\mathbf{y})G(\mathbf{x},\mathbf{y}) \dV(\mathbf{y})$ where $G$ is the known Green's function\footnote{For instance, in 2D $G(\mathbf{x},\mathbf{y})=\frac{i}{4} H_0(k |\mathbf{x}-\mathbf{y} |)$ where $H_0$ is the zero-order Hankel function of the first kind, see  \cite{Morse} for general formulas.}.

Setting $p^h=p-p^c$, the problem \eqref{equation: reference problem} can be rewritten in the following form:
\begin{equation}\label{equation: reference problem hom}
  \textrm{Find $p^h$ such that} \qquad
\left |
\begin{aligned}
  \Delta p^h + k^2 p^h &= 0 \quad &&\textrm{in $\Omega$} && \text{(a)}\\
  Z p^h +   \frac{\partial p^h}{\partial n} &=h_d =\tilde{h}_d - Z p^c -   \frac{\partial p^c}{\partial n}\quad &&\textrm{over $ \partial_Z \Omega$} && \text{(b)} \\
  p^h  &=p_d=\tilde{p}_d-p^c \quad &&\textrm{over $ \partial_p \Omega$} && \text{(c)} \\
 \frac{\partial p^h}{\partial n}  &= v_d =\tilde{v}_d-\frac{\partial p^c}{\partial n} \quad &&\textrm{over $ \partial_v \Omega$} && \text{(d)} \\
\end{aligned} \right.
\end{equation}

As one can see, the problem \eqref{equation: reference problem} with a source term, could be rewritten in an equivalent problem with modified boundary conditions and no source term. In the following we will consider a solution strategy to solve \eqref{equation: reference problem hom} and we omit superscript $h$

\smallskip

Many weak formulations of this system can be proposed, see \cite{Moiola2014} for a review, in this paper we investigate the VTCR which can be viewed as an indirect Trefftz method applied within a discontinuous-Galerkin framework.

\subsection{One-domain VTCR formulation}

Let us introduce the following space:
\begin{equation}
\begin{aligned}
\sH(\Omega) &= \left\{ u \in L^2(\Omega) / \nabla u \in L^2(\Omega)^d, \Delta u\in L^2(\Omega)\right\} \\
& = \left\{ u \in H^1(\Omega) / \nabla u \in H^{\text{div}}(\Omega)\right\} \\
\end{aligned}
\end{equation}
$\sH(\Omega)$ is an Hilbert space for the norm $\|u\|_\sH^2 = \|u\|_{H^1}^2+\|\Delta u\|^2_{L^2}$ \cite{dautraylions}. Assuming sufficient regularity on the shape of $\Omega$, the trace and normal flux are continuous operators on $\sH(\Omega)$ with values in $H^{\frac{1}{2}}(\partial\Omega)$ and $H^{-\frac{1}{2}}(\partial\Omega)$.

We consider the following subspace of $\sH(\Omega)$:
\begin{equation}\label{espace admissible }
  \sS(\Omega) = \left \{  p \in \sH(\Omega) \quad / \quad \Delta p + k^2 p =0 \textrm{\ in\ } \Omega     \right \}
\end{equation}
$\sS(\Omega)$ is a closed subspace of $\sH(\Omega)$ on which the {usual norm $\|u\|_{\sS}^2=\|\nabla u\|^2_{L^2}+|k|^2\|u\|^2_{L^2}$ is equivalent to the $\sH(\Omega)$ norm and to the $H^1(\Omega)$ norm (with equivalence coefficient dependent on $k$). Moreover $\sS(\Omega)$ is compactly embedded in $L^2(\Omega)$. 

In the following we make use of the $\sS$-norm and refer to it as the ``energy norm'', though the analysis is valid for any equivalent norm (which might be more pertinent from a physical point of view).
}

The general one-domain VTCR formulation of the problem consists in weakly enforcing the boundary conditions as follows ($\alpha\in \C$ is a parameter of the formulation):
\begin{equation}
\begin{aligned}
\text{find }p\in\sS(\Omega)\ & /\ \forall q\in\sS(\Omega),\ a(p,q)=l(q)\text{ with }\\
a(p,q)= &   \frac{1}{2} \int_{ \partial_Z \Omega} \left(  Z p + \frac{\partial p}{\partial n}  \right ) \left(  \alpha  \overline{q} + \frac{\overline{\alpha}}{Z}\frac{\partial \overline{q}}{\partial n}\right )  \ dS 
+ \overline{\alpha}\int_{ \partial_p \Omega}  p  \frac{\partial \overline{q}}{\partial n} \ dS + \alpha 
 \int_{ \partial_v \Omega}  \frac{\partial p}{\partial n}  \overline{q} dS  \\
l(q)= &   \frac{1}{2} \int_{ \partial_Z \Omega} h_d \left(  \alpha \overline{q} + \frac{\overline{\alpha}}{Z}\frac{\partial \overline{q}}{\partial n}\right )  \ dS 
 + \overline{\alpha}\int_{ \partial_p \Omega}  p_d  \frac{\partial \overline{q}}{\partial n} \ dS + \alpha 
 \int_{ \partial_v \Omega}  v_d  \overline{q} dS 
\end{aligned} \label{formulation}
\end{equation}
Clearly this formulation\footnote{{
To comply with the given subspace, the formulation should have been written with duality brackets in  $H^{1/2}$. The integral notation does not alter computations and it is correct for the fields used in practice which are the restriction of $C^\infty(\R^d)$ fields, assuming sufficiently regular loads (for instance $p_d\in H^{1/2}(\partial_p\Omega)$, $v_d\in L^2(\partial_v\Omega)$, $h_d\in L^2(\partial_Z\Omega)$).}} is consistent with the homogeneous reference system  in the sense that the solution of \eqref{equation: reference problem hom} satisfies the weak formulation. Also $a$~is a sesquilinear form, $l$~is antilinear, both are continuous by Cauchy-Schwarz inequality and the continuity of the trace and normal flux in $\sS(\Omega)$.

We have the following property:
\begin{equation}\label{eq:coerc1}
\begin{aligned}
\Real(a(p,p)) & = \frac{1}{2}\left( \Real(\alpha Z) \right)\int_{\partial_Z\Omega} \left(|p|^2+ \frac{1}{|Z|^2}\left|\frac{\partial p}{\partial n}\right|^2 \right) dS 
+ \Real\left(\alpha\int_{\partial\Omega} \frac{\partial p}{\partial n }\overline{p} dS \right) \\
&= \frac{1}{2}\left(\Real( \alpha Z)\right)\int_{\partial_Z\Omega} \left(|p|^2+ \frac{1}{|Z|^2}\left|\frac{\partial p}{\partial n}\right|^2 \right) dS \\
&\qquad+ \Real(\alpha) \|\nabla p\|^2_{L^2(\Omega)} - \Real(\alpha k^2)\|p\|^2_{L^2(\Omega)}  
\end{aligned}
\end{equation}
We now discuss the well posedness of the formulation (we recall that we supposed $\Real(Z)\geqslant 0$, $\eta\geqslant 0$):
\begin{itemize}
\item In the case of a negative imaginary part of the impedance and non-zero damping, the choice $\alpha=i\overline{k}$ gives a coercive formulation:
\begin{equation}
\left. \begin{aligned}
\Imag(Z)\leqslant 0\\ \eta>0 
\end{aligned}\right\}\ \alpha=i\overline{k} \text{ leads to }\Real(a(p,p))  \geqslant  \frac{\eta\omega}{c}\left(\|\nabla p\|^2_{L^2(\Omega)} +|k|^2\|p\|^2_{L^2(\Omega)}\right)
\end{equation}
\item {The domain of coercivity can be extended to impedance with small positive imaginary part if the system is sufficiently damped. Indeed, if there exists $\beta$ such that $\frac{\Real(Z)}{\Imag(Z)}\geqslant \beta> \frac{1-\eta^2}{2\eta}$ then setting $\alpha=(1+\rmi\beta)$  in \eqref{eq:coerc1} leads to: 
\begin{equation}
\begin{aligned}
\Real(a(p,p)) &= \|\nabla p\|^2_{L^2(\Omega)} + \frac{\Real(Z)-\beta\Imag(Z)}{2}\int_{\partial_Z\Omega} \left(|p|^2+ \frac{1}{|Z|^2}\left|\frac{\partial p}{\partial n}\right|^2 \right) dS\\&\qquad - (\frac{\omega^2}{c^2})(1-\eta^2-2\beta \eta)\|p\|^2_{L^2(\Omega)} \\
& \geqslant \|\nabla p\|^2_{L^2(\Omega)} +(\frac{\omega^2}{c^2})   \underset{> 0}{\underbrace{(\eta^2+2\beta \eta-1)}}\|p\|^2_{L^2(\Omega)}
\end{aligned}  
\end{equation}}
\end{itemize}
The coercivity implies existence, uniqueness, and continuity with respect to the loading of the solution, and the good convergence properties of Galerkin's approximations, in particular Cea's lemma applies.
\begin{itemize}
\item With the minimal assumptions $\Real(Z)\geqslant 0$, $\eta\geqslant 0$, the simple choice $\alpha=1$ gives a G\r{a}rding inequality:
\begin{equation}
\begin{aligned}
\alpha=1 \text{ leads to }\Real(a(p,p)) & \geqslant  \|\nabla p\|^2_{L^2(\Omega)} -\left(\frac{\omega}{c}\right)^2(1-\eta^2)\|p\|^2_{L^2(\Omega)}  
\end{aligned}
\end{equation}
\end{itemize}
The G\r{a}rding inequality (coercivity with respect to $L^2(\Omega)$) together with the compact embedding of $\sS(\Omega)$ in $L^2(\Omega)$ suffices to ensure existence, uniqueness and continuity with respect to the loading when $k^2$ is not an eigenvalue of the Laplacian with the given boundary conditions; the properties of Galerkin's approximations are ensured only for sufficiently large subspaces \cite{Moiola2014}.

\subsection{Many subdomains VTCR formulation}

The VTCR for many subdomains can be analysed in the framework of non-symmetric discontinuous Galerkin methods \cite{dg12}. 
Let $\parti$ be a partition of $\Omega$ into $N_{\Omega}$ non-overlapping sub-cavities $\Omega_E$ ($1\leqslant E\leqslant N_{\Omega}$), and let  $\Gamma_{E,E'} =  \partial \Omega_E \cap \partial \Omega_{E'}$ be  the face between subdomains $E$ and $E'$. Let $\setF$ be the set of faces, we arbitrarily attribute an orientation to each face.

In order to treat independently the fields in each subdomain, we introduce $\sST(\Omega)$ a broken version of $\sS(\Omega)$:
\begin{equation}
 \sST(\Omega) = \left\{ u\in L^2(\Omega)\ /\ \forall \Omega_E\in \parti,\ u_{|\Omega_E} \in \sS(\Omega_E) \right\}
\end{equation}
and we define the jump and average operators:
\begin{equation}
\forall\Gamma\in\setF, \Gamma=\partial \Omega_E \cap \partial \Omega_{E'},\begin{aligned}&
 \jump{ u}_\Gamma = u_{|\Omega_E} -  u_{|\Omega_{E'}} \\
 &  \mean {u} _\Gamma = \frac{u_{|\Omega_E} +  u_{|\Omega_{E'}}}{2}
 \end{aligned}
\end{equation}

We then have the following characterization of $\sS(\Omega)$:
\begin{equation}
u\in \sS(\Omega) \Leftrightarrow \left\{ u\in\sST(\Omega)/\ \forall \Gamma \in \setF,\ \jump{ u }_\Gamma =0,\ \mean{ \frac{\partial u}{\partial n}}_\Gamma =0 \right\}
\end{equation}
In other words, fields in $\sS(\Omega)$ satisfy continuity and balance of normal flux conditions on the interfaces:
\begin{equation*}
  p_E=p_{E'} \quad\text{ and }\quad
  \frac{\partial p_E}{\partial n_E} +\frac{\partial p_{E'}}{\partial n_{E'}}=0\quad\textrm{ over }\quad \Gamma_{E,E'}
\end{equation*}

In the VTCR, as in discontinuous Galerkin methods, the interface conditions are introduced inside the sesquilinear form. 
For $(p,q)\in \sST(\Omega)^2$, we note $a_E(p,q)=a(p_{|\Omega_E},q_{|\Omega_E})$ and $l_E(q)=l(q_{|\Omega_E})$. The VTCR for many subdomains writes:
\begin{equation}\label{eq:VTCRssd}
\begin{aligned}
\text{find }p\in\sST(\Omega)\ & /\ \forall q\in\sST(\Omega),\ a_\parti(p,q)=l_\parti(q)\text{ with }\\
a_\parti(p,q)= & \sum_E a_E(p,q) + \sum_{\Gamma\in\setF} \int_\Gamma \left(\jump{p}_\Gamma\mean{\frac{\partial \overline{q}}{\partial n}}_\Gamma - \jump{\overline{q}}_\Gamma\mean{\frac{\partial p}{\partial n}}_\Gamma\right)dS\\
l_\parti(q)= &  \sum_E l_E(q)
\end{aligned}
\end{equation}

Thus  the interface conditions are weakly imposed by an anti-hermitian formulation. This choice makes the broken formulation satisfies the same coercivity (or G\r{a}rding) inequality as the one-domain VTCR, so that no stabilization is required for the problem to be well posed. The same idea was used in Oden, Babuška and Bauman's discontinuous Galerkin formulation of the Poisson problem \cite{obb98}.

\subsection{Finite-dimensional approximation space}\label{discretization}
In order to build a finite dimension approximation subspace of $\sST(\Omega)$, we use the fact that on each subcavity $\Omega_E$ the pressure can be represented by \emph{Herglotz wave functions}. Let $\mathcal{C}$ be the unit sphere in $\R^d$ and  $x_E$ a reference point located in $\Omega_E$, we define Herglotz operator $\mathcal{H}$:
\begin{equation}\label{herglotz representation}
\begin{aligned}
\mathcal{H}_E : L^2(\mathcal{C}) &\to \sS(\Omega_E) \\
A &\mapsto \mathcal{H}_E[A]=p^E_A: x \mapsto\int_{\mathcal{C}} A(s)e^{\rmi k s\cdot(x-x_E)} ds
\end{aligned}
\end{equation}
Under regularity assumption on the shape of $\Omega_E$ (for instance $\partial \Omega_E$ is Lipschitz and $\Omega_E$ is star-shaped with respect to the point $x_E$), $\operatorname{range}(\mathcal{H}_E)$ is dense in $\sS(\Omega_E)$ \cite{colton01, Herglotz04} so that instead of searching $p\in\sST(\Omega)$ we can seek for $(A^E)\in L^2(\mathcal{C})^{N_\Omega}$. Keeping terminology adopted in previous work on the VTCR, the density $A^E$ is called the \emph{amplitude distribution} of $p_A^E$ since somehow the pressure is represented by a superposition of plane waves $e^{\rmi k s\cdot(x-x_E)}$ in direction $s$ with magnitude $A^E(s)$. The formulation in terms of unknown amplitudes makes the VTCR an indirect Trefftz method.
\smallskip

The retained strategy is then to discretize the space of amplitude distribution $L^2(\mathcal{C})^{N_\Omega}$. For a given subdomain $E$, the approximation subspace $\A\E$ of ${L^2(\mathcal{C})}$ of dimension $N_E$ is defined by a basis $\basA\E=(\ldots,A_n\E,\ldots)$. The resulting subspace of pressure is denoted $\sS^{N_E}(\Omega\E)$, it is spanned by $\basPA\E=\left(\ldots,\mathcal{H}_E(A_n\E),\ldots\right)$.
Various discretizations have been tested in previous implementations of the VTCR \cite{riou2008,kovalevsky11}. To keep expressions simple, we recall them  in the 2D case where $s$ only depends on one angle $\theta$: $s_\theta=\begin{pmatrix}\cos(\theta)\\\sin(\theta)\end{pmatrix}$.
\begin{itemize}
\item[$D_B$:]  $\basA\E$ is made out of piecewise-constant functions on $\mathcal{C}$, $\basPA\E$ is then a collection of wave Band functions: 
\begin{equation}
  \basPA\E = \left(\ldots,  \int_{\frac{2 \pi n}{N_E}}^{\frac{2 \pi (n+1)}{N_E}}    e^{\rmi k s_\theta \cdot(x-x_E)} \dth ,\ldots       \right)_{ 0 \leqslant n< N_E} \label{eq:Band}
\end{equation} 
\item[$D_F$:] $\basA\E$ is taken as a truncated Fourier series, $\basPA\E$ then being a collection of Fourier wave functions ($N_E$ is assumed to be odd): 
\begin{equation}
  \basPA\E =\left( \ldots, \int_{-\pi}^{\pi}  e^{\rmi n \theta}e^{\rmi ks_\theta \cdot(x-x_E)} \dth,\ldots  \right)_{ -\frac{N_E-1}{2} \leqslant n \leqslant \frac{N_E-1}{2}} \label{eq:Fourier}
\end{equation}
The main advantages of this discretization is that the functions of $\basPA\E$ can be evaluated analytically at any point $x$. Moreover these functions form a hierarchical basis, which makes it easy to increase the degree of the approximation, in particular a control of the condition number, similar to \cite{huttunen2002computational}, was proposed in \cite{kovalevsky11}.
\item[$D_\delta$:] One last possibility is to extend the definition of Herglotz {to a larger space containing Dirac distributions\footnote{For instance $H^{-s}(\mathcal{C})$ with $s>\frac{d-1}{2}$ where $d$ is the dimension of the physical space}. The typical choice is $\basA\E$ constituted by $N_E$ Dirac distributions supported at angular locations $\theta_n=\frac{2 \pi n}{N_E}$ (in some publications these functions are also called ``rays''). $\basPA\E$ is then the collection of a finite number of plane waves:
\begin{equation}
  \basPA\E = \left( \ldots,   e^{\rmi k s_{\theta_n}\cdot(x-x_E)},\ldots \right)_{0 \leqslant n< N_E} \label{eq:Dirac}
\end{equation}
This discretization presents the main advantage to allow analytical integration of the components of the weak formulation~\eqref{eq:VTCRssd} on straight lines, it also corresponds exactly to the enrichment functions used in the Discontinuous Enrichment Method \cite{farhatu}. Unfortunately it does not fit exactly the theory described in the following, which is why we propose an informal extension to that discretization in section~\ref{ssec:informalext}.}

\end{itemize}

The above discretizations, which have their pros and cons, lead to very similar asymptotic accuracy, as seen in comparisons presented in~\cite{riou2008,kovalevsky11}. From these studies, a priori criteria have emerged to choose a sufficiently fine discretization, represented by the dimension of the search space $N_E$.  In the Fourier case~\eqref{eq:Fourier}, $N_E$ can be set \emph{a priori} using an energetic criterion \cite{kovalevsky11}. In Dirac~\eqref{eq:Dirac} and Band~\eqref{eq:Band} cases, $N_E$ can be set \emph{a priori} using a geometrical heuristic criterion \begin{equation} \label{criterion}
N_E= \operatorname{round}(\mu \pi R_E/\lambda)
\end{equation}
where $R_E$ is the characteristic diameter of $\Omega_E$, $\lambda$ is the wavelength and $\mu$ a positive real number close to 1 (see \cite{desmetu2}  eqn. 43, or \cite{sourcis} eqn. 3.37, and the example of section \ref{subsection: scattering of a cylinder}).

\subsection{Discrete system}
\newcommand{\bK}{\mathbf{K}}
\newcommand{\bF}{\mathbf{f}}
Introducing any of the previous basis in the weak formulation \eqref{eq:VTCRssd} leads to a linear system with the following block structure:
\begin{equation}\label{eq:syslin}
\bK \bal = \bF, \text{ with }
\bK=\begin{pmatrix}
\bK_{11} & \bK_{12} & &\ldots & & \bK_{1 N_\Omega} \\
 -\bK^H_{12} & \bK_{22} & & & & \\
 \vdots  &  & &\ddots & & \\
 -\bK^H_{1 N_\Omega} &  & &  \ldots && \bK_{N_\Omega N_\Omega}\\
\end{pmatrix}, \bal=\begin{pmatrix}
\bal_1 \\ \vdots \\ \\ \bal_{N_\Omega}
\end{pmatrix}, \bF=\begin{pmatrix}
\bF_1 \\ \vdots \\ \\ \bF_{N_\Omega}
\end{pmatrix}
\end{equation}
where exponent $H$ stands for the conjugation-transposition, and $\bal_E$ corresponds to the vector of unknown amplitude of the chosen basis functions in the subdomain $\Omega_E$. 
The off-diagonal blocks correspond to the coupling between subdomains, they are zero for non-neighbors subdomains so that the system has a sparse-by-block structure. In the case where damping ensures coercivity of the formulation, the diagonal blocks are positive. 

Each degree of freedom is associated with one wave function in one subdomain and it necessarily contributes to the off-diagonal coupling term. This implies that contrary to what is commonly encountered in DG methods \cite{doi:10.1137/070706616,Nguyen20093232} there is no ``internal'' degree of freedom and condensation can not apply.

\section{The Optimized space of approximation}\label{OMG}

One major problem with searching for amplitude components is that operator $\mathcal{H}_E$ is compact in $L^2(\mathcal{C})$: there is an accumulation of eigenvalues near zero. In other words there exist amplitudes of unit norm in $L^2(\mathcal{C})$ capable to create arbitrary small pressure fields in $\sS(\Omega_E)$: 
\begin{equation}
\forall \varepsilon >0,\ \exists A\in L^2(\mathcal{C}) \text{ with }\|A\|_{L^2(\mathcal{C})}=1 \text{ and } \|\mathcal{H}_E[A]\|_{\sS(\Omega_E)}<\varepsilon
\end{equation} 
To illustrate the consequences of this problem, let us consider the favorable case of a coercive formulation in $\mathcal{S}^\mathcal{T}(\Omega)$ where there exists a positive constant $Q$ such that:
\begin{equation}
\forall p\in \mathcal{S}^\mathcal{T}(\Omega),\ a_\mathcal{T}(p,p) \geqslant Q \|p\|^2_{\sS(\Omega)}
\end{equation}
If we now consider the formulation in terms of amplitudes, all we can say is:
\begin{equation}\label{eq:losscoerc}
\forall (A\E)\in L^2(\mathcal{C})^{N_\Omega},\ a_\mathcal{T}((\mathcal{H}_E[A\E]),(\mathcal{H}_E[A\E])) \geqslant 0   %\mathbf{0} \sum_E \|A\E\|_{L^2(\mathcal{C})}^2
\end{equation}
the sesquilinear form is positive but it can not be bounded from below by  the norm of the amplitudes $\sum_E \|A\E\|_{L^2(\mathcal{C})}^2$, the formulation in amplitude is not coercive. Moreover, the discrete system is likely to be poorly conditioned: as soon as the discretization space is large enough, it is probable that it contains eigenvectors associated with small eigenvalues.

In this section, we propose to build a subspace which does not excite the near-zero eigenvalues; by construction, the coercivity is preserved in that subspace and the condition number is controlled. We also prove that using this subspace only leads to a small loss of precision. By analogy with domain decomposition or multigrid methods, the subspace will often be referred to as the coarse space of approximation.

%As said earlier one major problem of the VTCR is that the problem set in terms of amplitudes is not coercive since large amplitude can result in  negligible pressures without effect on the sesquilinear form.

\subsection{Notations}
The analysis we develop {is suited for discretizations in $L^2(\mathcal{C})$, either the Band discretization~\eqref{eq:Band} or the Fourier discretization~\eqref{eq:Fourier}. An informal extension to the Dirac discretization~\eqref{eq:Dirac} is proposed in section~\ref{ssec:informalext}.

Note that the analysis } can be conducted independently on each subdomain. For a given basis in the amplitude domain $\basA\E$, we write $\bal\E$ the vector of components of the amplitude $A\E$, which is associated with the pressure field~$p\E$:
\begin{equation*}
\begin{aligned}
s\in\mathcal{C}&\mapsto A\E(s)=\basA\E(s)\bal\E\\
x\in\Omega\E&\mapsto p\E(x)=\basPA\E(x)\bal\E
\end{aligned}
\end{equation*}

For the discretizations by piecewise constant amplitudes or by truncated Fourier series we note $\|\|_\mathcal{C}$ the $L^2(\mathcal{C})$-norm of amplitude vectors and $\massC\E$ mass matrix associated with basis $\basA\E$:
\begin{equation}
\|\bal^E\|_\mathcal{C}^2 = \|\basA\E\bal\E\|_{L^2(\mathcal{C})}^2 = \bal{\ET} \massC\E \bal\E 
\end{equation}
Note that for any of the proposed discretization methods, the mass matrix $\massC^E$ is diagonal.

\subsection{Principle}
We assume that for each subdomain $\Omega_E$, an approximation subspace was defined, and that, based on engineering rules presented in equation~\eqref{criterion}, it is sufficiently large for the solution to be approximated with enough accuracy.

{We propose to select a smaller approximation subspace where amplitude generate non-negligible pressure. Note that this selection can be conducted independently on each subdomain, in parallel.
}
For any subdomain $\Omega_E$, we would like to select {the largest}  subspace $\At_\sigma^E$ of $\A^E$ where the following property is verified:
\begin{equation}\label{eq:defst}
\forall A \in \At_\sigma^E,\ \|\mathcal{H}_E[A]\|_{\sS(\Omega_E)} \geqslant \sigma \| A \|_{L^2(\mathcal{C})}
\end{equation}
where $\sigma>0$ is a parameter to be defined by the user. In that subspace, the equation~\eqref{eq:losscoerc} becomes:
\begin{equation}\label{eq:losscoerc2}
\forall (A\E)\in \prod(\At_\sigma^E),\ a_\mathcal{T}((\mathcal{H}_E[A\E]),(\mathcal{H}_E[A\E])) \geqslant \sigma^2 Q \left(\sum_E \|A\E\|_{L^2(\mathcal{C})}^2\right)
\end{equation}
In other words, coercivity (and condition number) is controlled by $\sigma$ . The smaller $\sigma$, the larger $\At_\sigma^E$ and the poorer the condition number is.

\newcommand{\bpsi}{\boldsymbol{\psi}}
\subsection{Construction of the subspace}\label{ssec:construct}
The subspace $\At_\sigma\E$ is not practical to compute. We thus propose to a way to approximate it. The idea is to use the property of finite element interpolation to obtain a good estimation of the $\sS$ norm. Let us consider a mesh $T\E_h$ of $\Omega_E$. Let $({x}_i\E)_{1\leqslant i\leqslant N\E_h}$ denote the nodes of $T\E_h$ and $\bpsi\E=(\ldots,\psi_i\E,\ldots)$ be associated matrix of shape functions (of degree $q$). $h$ stands for the maximal length of the edges of the mesh, it must be chosen in agreement with the characteristic length of the problem $\Real\left(\frac{1}{k}\right)$. Anyhow this mesh is used to make interpolations and compute norms, not to approximate the solution so that it can be much coarser than recommended for finite element computations. Moreover the meshes $(T\E_h)_E$ do not need to be compatible at interfaces ($h$ could even be set independently on each subdomain).

Let $\basPAh\E$ be the matrix of the pressure field generated by the basis $\basA\E$ evaluated at the nodes of the mesh:
\begin{equation}
\basPAh\E = \begin{pmatrix} 
& \vdots & \\
\ldots & \mathcal{H}_E[A_j\E](x_i\E) & \ldots \\
& \vdots & 
\end{pmatrix},\qquad \text{row } i,\text{ column }j
\end{equation}
A field of $\mathcal{S}^{N_E}(\Omega_E)$ writes $\basPA\E\bal\E$, its finite element interpolation is $\bpsi\E\basPAh\E\bal\E$, the classical properties of finite element interpolation ensure that the distance between these two fields  can be controlled:  there exist a constant $C\E$ which depends on the mesh and on $k$, and which can be decreased at will by decreasing the characteristic size $h$ of the mesh or increasing the degree $q$ of the interpolation~\cite{brenner}, such that:
\begin{equation}
\forall \bal\E,\qquad \|\basPA\E\bal\E-\bpsi\E\basPAh\E\bal\E\|_{\sS(\Omega_E)}   \leqslant C\E \|\bal^E\|_\mathcal{C}
\end{equation}

Let $\massh\E$ be the $\sS(\Omega\E)$-mass matrix associated with the mesh $T\E_h$, we have:
\begin{equation}
\begin{aligned}
\|\bpsi\E\basPAh\E\bal\E\|^2_{\sS(\Omega\E)} &= \bal{\ET}\basPAh\ET \massh\E\basPAh\E \bal\E \\
\end{aligned}
\end{equation}

The approximation $\At_{\sigma,h}\E$ of $\At_\sigma\E$ defined in \eqref{eq:defst} is generated by all the combinations of amplitudes which create interpolated pressure fields of sufficient norm. In other words, we solve the following problem: 
\begin{equation}
\text{find }\bal\E\in\C^{N_E}/\ \bal{\ET}\basPAh\ET \massh\E\basPAh\E \bal\E  \geqslant \sigma^2 \bal{\ET} \massC\E \bal\E
\end{equation}
This corresponds to the selection of eigenvectors associated with generalized eigenvalues larger than $\sigma^2$ for the system of hermitian positive definite matrices $(\basPAh\ET \massh\E\basPAh\E ,\massC\E )$.

Let $\tV\E$ be the {subset} of $\massC\E$-normalized eigenvectors {${\bf V}\E$} associated with eigenvalues larger than $\sigma^2$, $\basA\E\tV\E$ is a basis of $\At\E_{\sigma,h}$. If we assume that the mesh was built such that $\sigma>C\E$, we have:
\begin{equation}
\begin{aligned}
\|\basPA\E\tV\E\tilde{\bal}\E\|_{\sS(\Omega_E)}& = \|\bpsi\E\basPAh\E\tV\E\tilde{\bal}\E + \left(\basPA\E\tV\E\tilde{\bal}\E-\bpsi\E\basPAh\E\tV\E\tilde{\bal}\E\right)  \|_{\sS(\Omega_E)}  \\
& \geqslant \left(\sigma -C\E\right) \|\tilde{\bal}\E\|_{\mathcal{C}}
\end{aligned}
\end{equation}
This inequality shows that the use of a mesh to estimate the $\sS$-norm does not prevent to control coercivity \eqref{eq:losscoerc2} as long as the mesh is fine enough for constant $C\E$ to be small with respect to $\sigma$. Indeed the coercivity relation becomes:
\begin{equation}
\forall (\tilde{\bal}\E)\in \prod \C^{\tilde{N}_E}, \ a_\mathcal{T}((\basPA\E\tV\E\tilde{\bal}\E),(\basPA\E\tV\E\tilde{\bal}\E)) \geqslant (\sigma-\max_E C^E)^2 Q \left(\sum_E \|\tilde{\bal}\E\|^2_{\mathcal{C}}\right)
\end{equation}

\subsection{Loss of precision}

Searching the solution in the subspace $\At\E_{\sigma,h}$ instead of $\A\E$, one expects a loss of precision. We thus need to estimate the ability of $\At\E_{\sigma,h}$ to approximate fields in $\A\E$.
 For any given components $\bal\E$ of a vector { in the approximation subspace $\A\E$}, let $\tilde{\bal}\E=\tV\E\tV\ET\bal\E$ be its $\massC\E$-orthogonal projection on {the subspace $\At\E_{\sigma,h}$}, we have:
\begin{equation}
\begin{aligned}
\|\basPA\E\bal\E - \basPA\E\tilde{\bal}\E\|_{\sS(\Omega\E)}  &\leqslant \|\basPA\E\bal\E - \bpsi\E\basPAh\E\bal\E \|_{\sS(\Omega\E)} \\ &\quad+\|\bpsi\E\basPAh\E\bal\E - \bpsi\E\basPAh\E\tilde{\bal}\E)\|_{\sS(\Omega\E)} \\&\quad+\|(\bpsi\E\basPAh\E\tilde{\bal}\E - \basPA\E\tilde{\bal}\E)\|_{\sS(\Omega\E)} \\
&\leqslant C\E (\|\bal\E\|_{\mathcal{C}} + \|\tilde{\bal}\E\|_{\mathcal{C}}) + \sigma \|\bal\E\|_{\mathcal{C}}\\
&\leqslant \left(2C\E + \sigma \right)\|\bal\E\|_{\mathcal{C}} \leqslant 3\sigma\|\bal\E\|_{\mathcal{C}} 
\end{aligned}
\end{equation}
The extra error due to the truncation is thus directly controlled by $\sigma$ (for a mesh satisfying $\sigma>C\E$).

{\subsection{Informal extension for the discretization with Dirac's distributions}\label{ssec:informalext}

The discretization of the amplitude space by Dirac distributions~\eqref{eq:Dirac} can not be conducted in $L^2(\mathcal{C})$. We extend the method to that case by arbitrarily setting $\massC^E=\mathbf{I}$; the (improper) notation $\|A^E\|_{L^2(\mathcal{C})}$ then stands for the Euclidean norm of the components $\|\bal^E\|_2$. 
}

\section{Practical considerations}\label{sec:practical}

\subsection{Setting of the threshold \texorpdfstring{$\sigma$}{Sigma}}
As seen earlier, the parameter $\sigma$ controls the coercivity (and then the condition number) and the attainable precision of the coarse problem set in $\At_\sigma\E$: the smaller $\sigma$ the larger the coarse problem, the poorer its condition number and the greater the precision of its solution.\medskip

We propose to choose $\sigma$ by an energy criterion: it is indeed possible to choose which fraction of the total energy is present in the coarse model.

For any of the choices of discretization, the basis $\basA\E$, the mass-matrix for the amplitudes $\massC\E$ is diagonal and easy to compute, we can at no extra cost transform the generalized eigenvalue system into a classical eigenvalue problem. The sum of the eigenvalues $(\theta_i)$ is then the trace of the matrix.

The criterion is then defined by a scalar $0<\beta\leqslant 1$, so that $(1-\beta)$ corresponds to the fraction of energy inside the coarse model:
\begin{equation}
\sigma \text{ such that }\sum_{\theta_i>\sigma}(\theta_i) > (1-\beta) \sum_{i=1}^{N_E} \theta_i = (1-\beta) \operatorname{trace}( (\massC\E)^{-\frac{1}{2}} \basPAh\ET \massh\E\basPAh\E (\massC\E)^{-\frac{1}{2}} )
\label{beta}
\end{equation}

In the examples,  we used the following values for $\beta=0.25, 0.1, 10^{-4},10^{-6}$ .

\subsection{Simplified evaluation of the \texorpdfstring{$\sS$}{S}-norm}

Regarding the evaluation of the $\sS(\Omega\E)$ norm, if the mesh $T\E_h$ is sufficiently fine and regular (no region is over or under meshed), then the condition number of $\massh\E$ (for the Euclidean norm) is $O(1)$ \cite{Kamenski14}, which means that the $\massh\E$-norm is correctly approximated by the Euclidean norm. In that case, the generalized eigenvalue problem becomes the singular values decomposition of the matrix $\basPAh\E (\massC\E)^{-\frac{1}{2}}$. 
The sum of all singular values {being equal to the (easy to compute)} Frobenius norm of matrix $\basPAh\E(\massC\E)^{-\frac{1}{2}}$, a criterion similar to the one described in previous subsection can then be applied to define $\sigma$.\medskip

In the following examples, we use discretization in {Dirac's (equation~\eqref{eq:Dirac} and subsection~\ref{ssec:informalext}) and the simplified evaluation of the $\sS$-norm, which is that case corresponds to computing the SVD of $\basPAh\E$ (since $\massC\E$ is set to identity). Unless stated otherwise, we use regular meshes with a characteristic size $h=\lambda /3$, where $\lambda$ is the wave-length. 
}

\subsection{Solver}

The coarse subspace $(\At_{\sigma,h}\E)$ can be directly used to find an approximate solution of the problem. Since the associated system  is well conditioned and small, a direct solver can be employed, moreover we proved that the added error is limited. It can also be used as the coarse space of a multigrid method or of an augmented solver. 

Let us briefly present how the subspace is used as an augmentation space for the LSQR Krylov solver \cite{lsqr82} which is known to behave correctly on VTCR systems. 
We introduce the basis of the coarse subspace, obtained by the diagonal concatenation of the subdomains' basis:
\begin{equation}
\tV=\begin{pmatrix}
\tV^1 & 0 &  \\
0 & \ddots & \\
& & \tV^{N_\Omega}
\end{pmatrix}
\end{equation}

LSQR solves for the normal equation:
\begin{equation}
\bK^H\bK \bal = \bK^H \bF
\end{equation} 
In that case, augmentation consists in ensuring that the residual $(\bF-\bK \bal)$ is orthogonal to $\operatorname{range}(\bK\tV)$. This is classically implemented by an initialization/projection $(\bal_0,\proj)$ method:
one sets $\bal=\proj\hat{\bal}+\bal_0$ and solves the system in $\hat{\bal}$ with a classical LSQR:
\begin{equation}
(\bK \proj) \hat{\bal} = \bF - \bK \bal_0
\end{equation}
where
\begin{equation}
\begin{aligned}
\bal_0& = \tV (\tV^H \bK^H \bK \tV)^{-1}\tV^H \bK^H \bF \\
\proj &=  \mathbf{I}-\tV (\tV^H \bK^H \bK \tV)^{-1}\tV^H \bK^H \bK
\end{aligned}
\end{equation}
Note that the coarse matrix $(\tV^H \bK^H \bK \tV)$ is coarse hermitian positive definite.
Using this method, one expects a good initialization (which makes the initial residual small). Note that because the coarse problem controls the higher part of the spectrum whereas the lower part is very populated near zero, the rate of convergence can not be much improved by the augmentation.

In the following the direct solution in the {coarse} subspace is referred to as O-VTCR, the augmented LSQR is A-VTCR. Of course, O-VTCR corresponds to the initialization of A-VTCR.

\section{Academic example: Scattering by a sound-hard cylinder} \label{subsection: scattering of a cylinder}

We first evaluate our method on an academic problem for which the analytical solution is known: the scattering of a plane wave by a sound-hard cylinder obstacle. 
We evaluate the classical VTCR equipped with various solvers: Matlab's direct solver ``$\backslash$'' (which in that case corresponds to LU-solver), Moore-Penrose pseudo inverse (Matlab's ``pinv''); and we compare them to the new reduced approach used either directly  with  Matlab's solver ``$\backslash$'' (O-VTCR, which in that case corresponds to a Cholesky solver) or used as the augmented LSQR approach (A-VTCR), both for $\beta$ of equation~\eqref{beta} equal to $\beta=0.25, 0.1, 10^{-4},10^{-6}$.  We compare the characteristics of the systems (size, condition number, attainable precision) as well as the convergence of the iterative solvers.
\smallskip

The sound-hard cylinder obstacle $S_1$ has a radius $R_1=0.5$m.  The surrounding acoustic medium $\Omega$ is considered to be air  ($\rho$ = 1.25 kg.m$^{-3}$, $c$ = 330 m.s$^{-1}$, and $\eta=0$). It is truncated by a concentric circular surface $S_2$ of radius $R_2=6$m. The problem \eqref{equation: reference problem} is solved with the following boundary conditions: 

\begin{equation}\label{equation: cylinder problem}
\left |
\begin{aligned}
\frac{\partial p}{\partial n}+\frac{\partial p_{scat}}{\partial n}  &=0   \quad &&\textrm{over $ S_1$} && \text{(a)} \\
  p + \imath \frac{c}{\omega}  \frac{\partial p}{\partial n} &= 0 \quad &&\textrm{over $ S_2$} && \text{(b)}  
\end{aligned} \right.
\end{equation}
where  $p_{\text{scat}} $ denotes a plane wave propagating in the direction $\theta = \pi$ and $\omega$ the circular frequency in rad.s$^{-1}$.
Equation \eqref{equation: cylinder problem}(b) is an absorbing condition that approximates the Sommerfeld radiation condition.

\begin{figure}
\centering
  \includegraphics[width=1\linewidth]{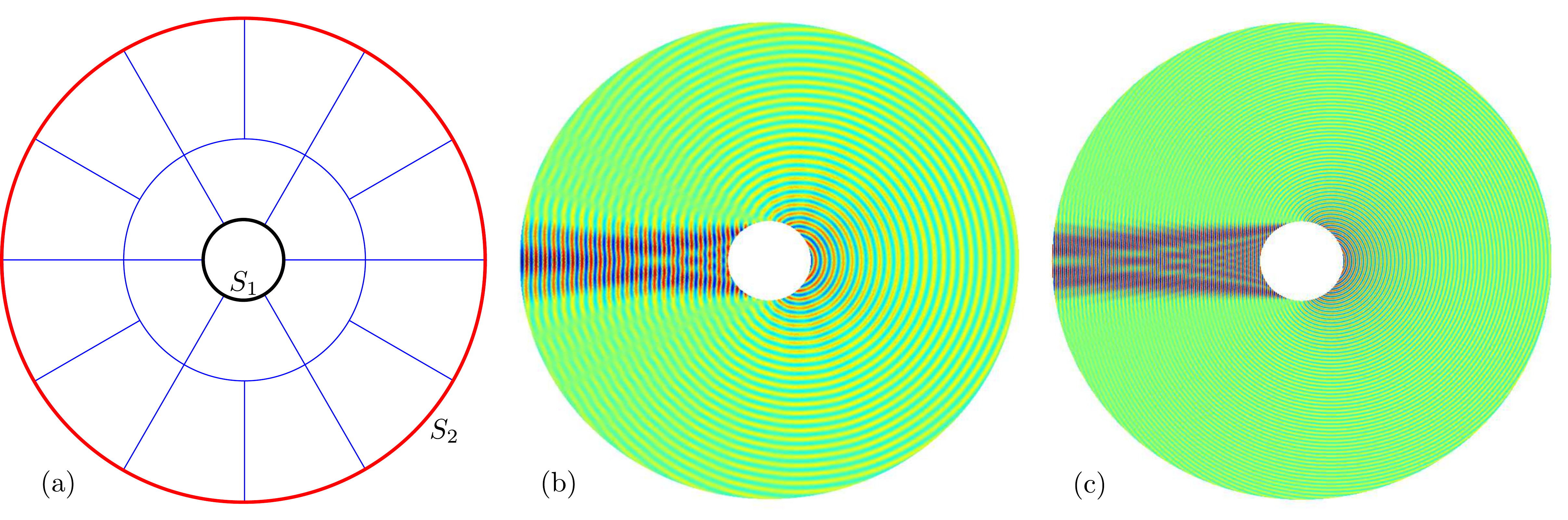} 
   \caption{Scattering by a sound-hard cylinder (Section \ref{subsection: scattering of a cylinder}): (a) discretized computational domain, (b) real part of the exact solution at f=1500Hz, (c) real part of the  exact solution at f=3500Hz .}
   \label{decompodisk}
\end{figure}

The exact solution of the resulting problem is known to have the  analytical form:
\begin{equation}
p_{ex}(r,\theta)=\sum_{m=0}^{\infty} \frac{\imath^m}{2-\delta_{0m}} \left( \frac{\text{J}_{m+1}(k R_1)-\text{J}_{m-1}(k R_1)}{\text{H}^{(2)}_{m+1}(k R_1)-\text{H}^{(2)}_{m-1}(k R_1)}   \right) \cos (\theta) \text{H}^{(2)}_{m}(k r)
\end{equation}
where $\delta_{0m}$ is the Kronecker's delta ($\delta_{0m}=1$ if $m=0$ and $\delta_{0m}=0$ otherwise), $\text{J}_{m}$ is the order $m$ Bessel function of the first kind, and $\text{H}^{(2)}_{m}$ is the order $m$ Hankel function of second kind.
Two different frequencies are considered f=1500 Hz and f=3500 Hz, corresponding respectively to $9$ and $21$ wavelengths in the diameter of the obstacle. The exact solutions are represented in Figure~\ref{decompodisk}. 

The domain $\Omega$ is decomposed into 18 sub-cavities as illustrated in Figure~\ref{decompodisk}.

Errors $\varepsilon[p]$ are evaluated using the following expression which incorporates contributions from pressure discontinuities across interfaces:
\begin{equation}\label{p:error:E}
\varepsilon[p]
 := \frac{1}{\|p_{ex}\|_{\sS(\Omega)}} \; \Biggl[ \;
   \|p-p_{ex}\|^{2}_{\sS(\Omega)} + \sum_{\Gamma\in\mathcal{F}} \|\jump{p}\|^{2}_{L^2(\Gamma)} \;
  \Biggr]^{1/2}
\end{equation}

We use the discretization by uniform Dirac's distribution (``rays'') \eqref{eq:Dirac}. 
In order to study the influence of the initial refinement of the discretization of the amplitude space, the number of rays in the subdomains $N_E=\operatorname{dim}(\A\E)$, is set using the criterion (\ref{criterion}) where the parameter $\mu$ varies from $0.1$ (insufficient discretization) to $2$ (a priori more than enough rays to represent the solution). The integrals needed in the calculation of the matrix $\mathbf{K}$ are performed with a numerical quadrature 
using a point density corresponding to 30 points per wavelength.

\subsection{Characteristics of the coarse model}

\begin{figure}[ht]
\centering
  \includegraphics[width=1\linewidth]{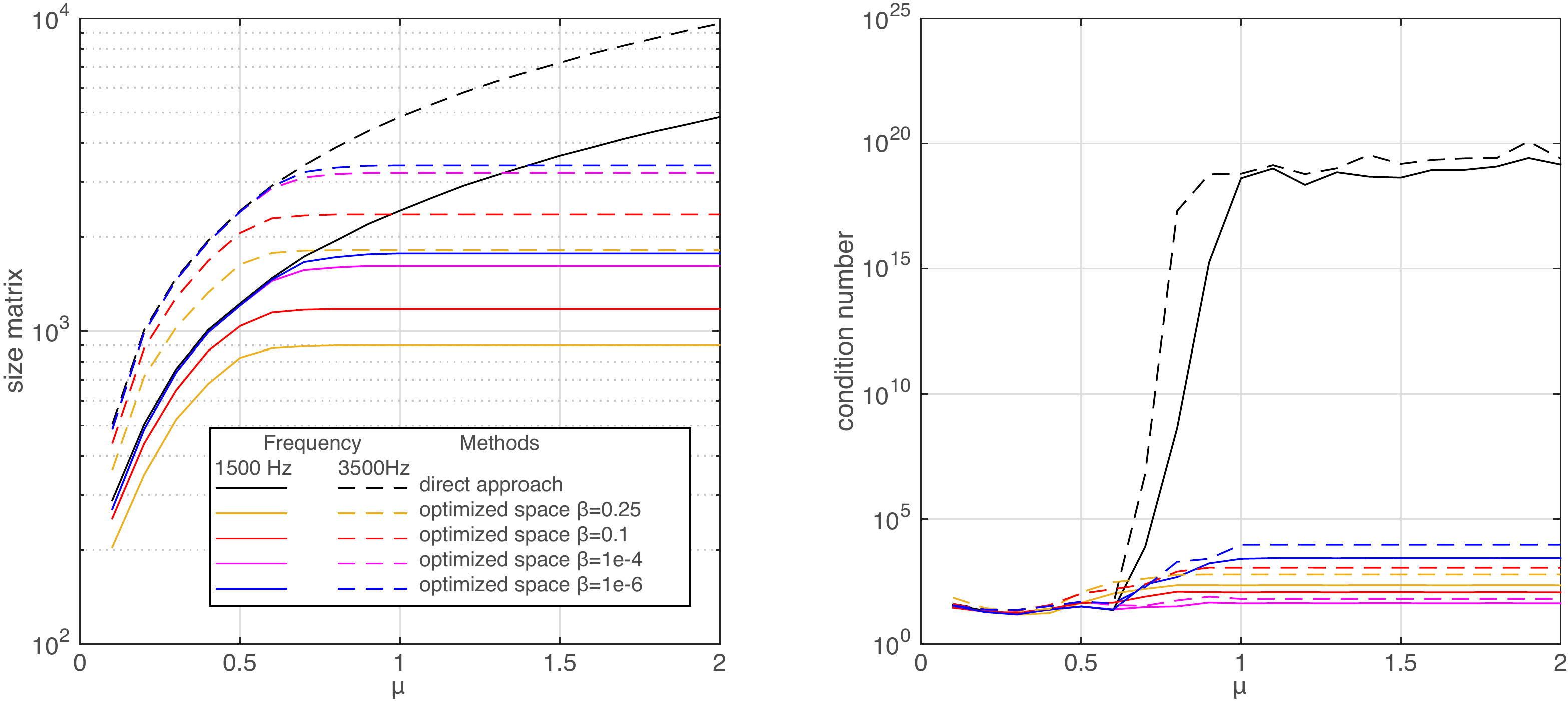} 
   \caption{Scattering by a sound-hard cylinder (Section \ref{subsection: scattering of a cylinder}): Evolution of the matrix size (left) and its condition number (right) with the parameter $\mu$ at the f=1500Hz (plain line) and f=3500Hz (dashed line) when using a direct VTCR approach (black) or the O-VTCR with $\beta=0.25, 0.1, 10^{-4},$ and $10^{-6}$ (respectively orange, red, magenta and blue).}
   \label{dofcondi}
\end{figure}

Figure~\ref{dofcondi} (left) presents the evolution of the size of the problems and (right) the evolution of the condition number as functions of the underlying discretization $\mu$ of the amplitude space, at frequencies $1\, 500$ Hz (plain curves) and $3\, 500$ Hz (dashed curves). 

The dimension of the approximation space $(\A\E)$ is linear in $\mu$ (see~\eqref{criterion}). The optimized subspaces $(\At\E)$ almost coincide with $(\A\E)$ for small $\mu$  but they tend not to grow for $\mu>0.75$: beyond this limit, increasing the size of the approximation space does not mean that more energy is present (in terms of pressure). This can in particular be observed by the small number of extra vectors required to capture a proportion of $(1-10^{-6})$ of the energy instead of  $(1-10^{-4})$. 

The condition number of the unfiltered system explodes around $\mu=0.7$. This also corresponds to the existence of almost zero energy vectors in the approximation subspace. On the contrary, the optimized subspace experience only a slight increase of the condition number which reaches at worse $10^4$.

\subsection{Attainable precision by the coarse model}
\begin{figure}[ht]
\centering
  \includegraphics[width=.9\linewidth]{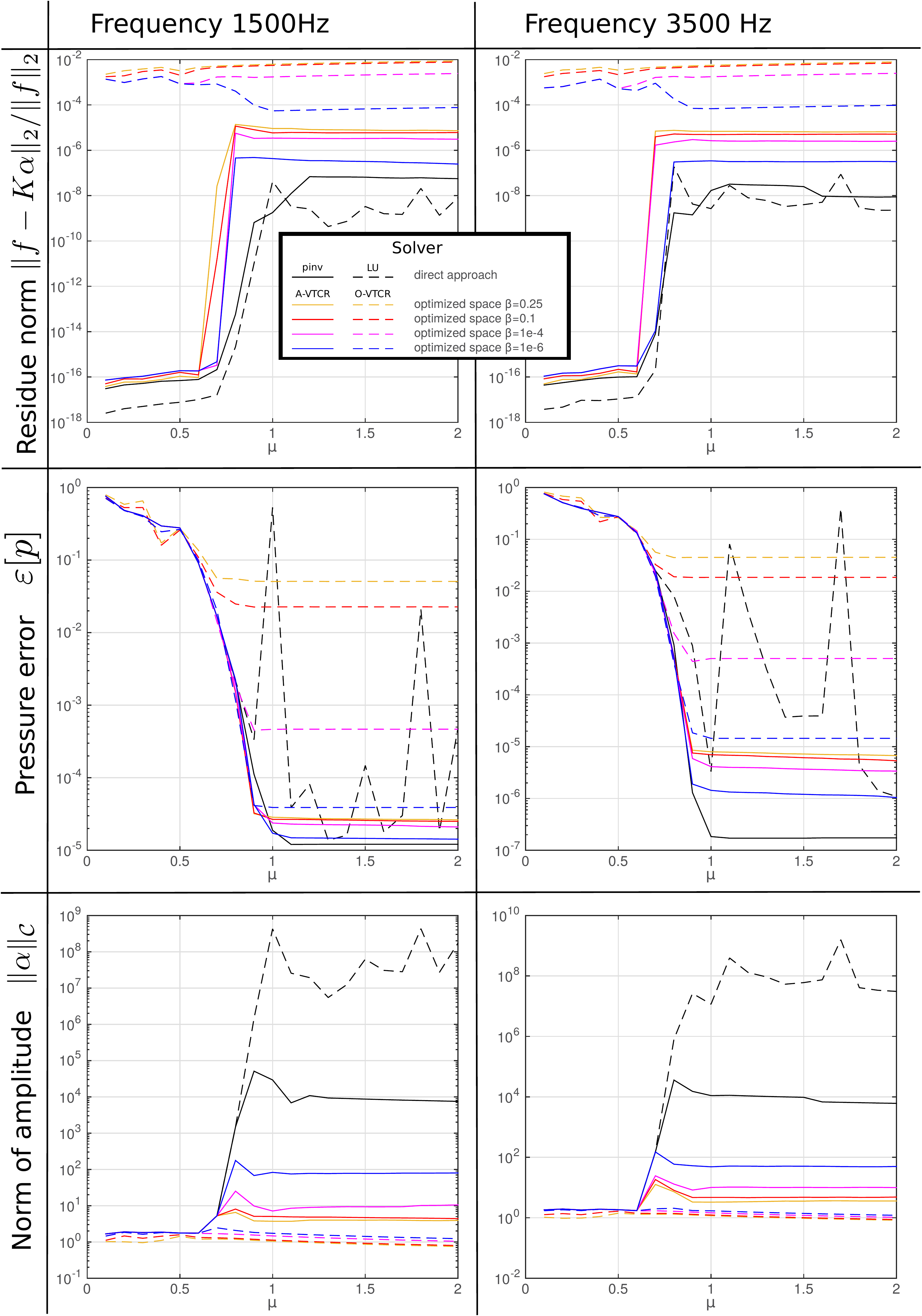} 
   \caption{Scattering by a sound-hard cylinder (Section \ref{subsection: scattering of a cylinder}): Evolution of the residual, pressure error and solution's norm with the discretization refinement $(\mu)$ at frequencies  f=1500Hz (left) and f=3500Hz (right) for the direct VTCR approach (black) or the A-VTCR and O-VTCR with $\beta=0.25, 0.1, 10^{-4}$ and $10^{-6}$ (respectively orange, red, magenta and blue).}
   \label{errorall}
\end{figure}

Figure~\ref{errorall} presents the residual, the pressure error (with respect to the analytical solution) and the norm of the solution, obtained by the solvers depending on the initial discretization (at frequencies 1\,500 Hz and 3\,500 Hz). The dashed lines correspond to the use of a direct solver on the complete (black) or on the reduced models (O-VTCR for various $\beta$), the plain lines correspond either to a pseudo inverse on the full model or to an augmented LSQR-solver (A-VTCR). For that study the number of LSQR iterations is equal to the dimension of the search space; this choice is meant to illustrate the bad conditioning of the system since the attained residual is far from unit round-off. Setting the stopping criterion for augmented-LSQR is discussed in next subsection.

For insufficient discretization ($\mu<0.7$), the {coarse} spaces practically coincide with the full space and condition number is low so that all solvers give comparable results (similar error in pressure, very small residual). 

After a threshold (which depends on $\mu$ and $\beta$), the {coarse} spaces do not evolve, and the associated performance of reduced models (O-VTCR) does not improve: for instance, at $1\,500$ Hz and for $\mu>0.9$,  $\beta=10^{-4}$ gives a precision $\varepsilon[p]$ of $5\, 10^{-4}$ and $\beta=10^{-6}$ gives a precision of $5\,10^{-5}$. Because of the better conditioning of the system, the precisions of the coarse models are much more stable than the precision of the full model with the same solver (black dashed curve). 

We observe that the correlation between the residual and the error strongly depends on $\mu$; and that for sufficient discretizations, precision is often much better than the residual can tell (the residual is hardly different between $\beta=10^{-1}$ and $\beta=10^{-4}$ whereas the pressure error is 15 times smaller). 

The norm of the amplitudes vector $\bal$ explodes with direct solvers when the discretization is sufficiently fine for good precision to be attained, whereas the O-VTCR gives an amplitude vector of almost constant norm (independent of $\mu$).

If a reduced O-VTCR model gives insufficient precision, it can be improved either by choosing $\beta$ closer to zero or by doing augmented LSQR iterations (A-VTCR). The obtained precision is presented by plain curves. Whatever $\beta$, the final precision of the pressure field is comparable with the one obtained by the reference pseudo-inverse solver. Here again the residual is not an excellent indicator for the quality of the pressure solution. We observe that the solutions given by the coarse models have very small amplitudes and that doing iterations slightly increases that norm but anyhow the result remains much smaller than what is obtained by the Moore-Penrose pseudo-inverse which is supposed to give the minimal norm solution. In other words, the pseudo-inverse's slightly smaller residual is paid by a strong increase of the amplitudes.

\subsection{Performance of the augmented solver}
\begin{figure}[ht]
\centering
  \includegraphics[width=1\linewidth]{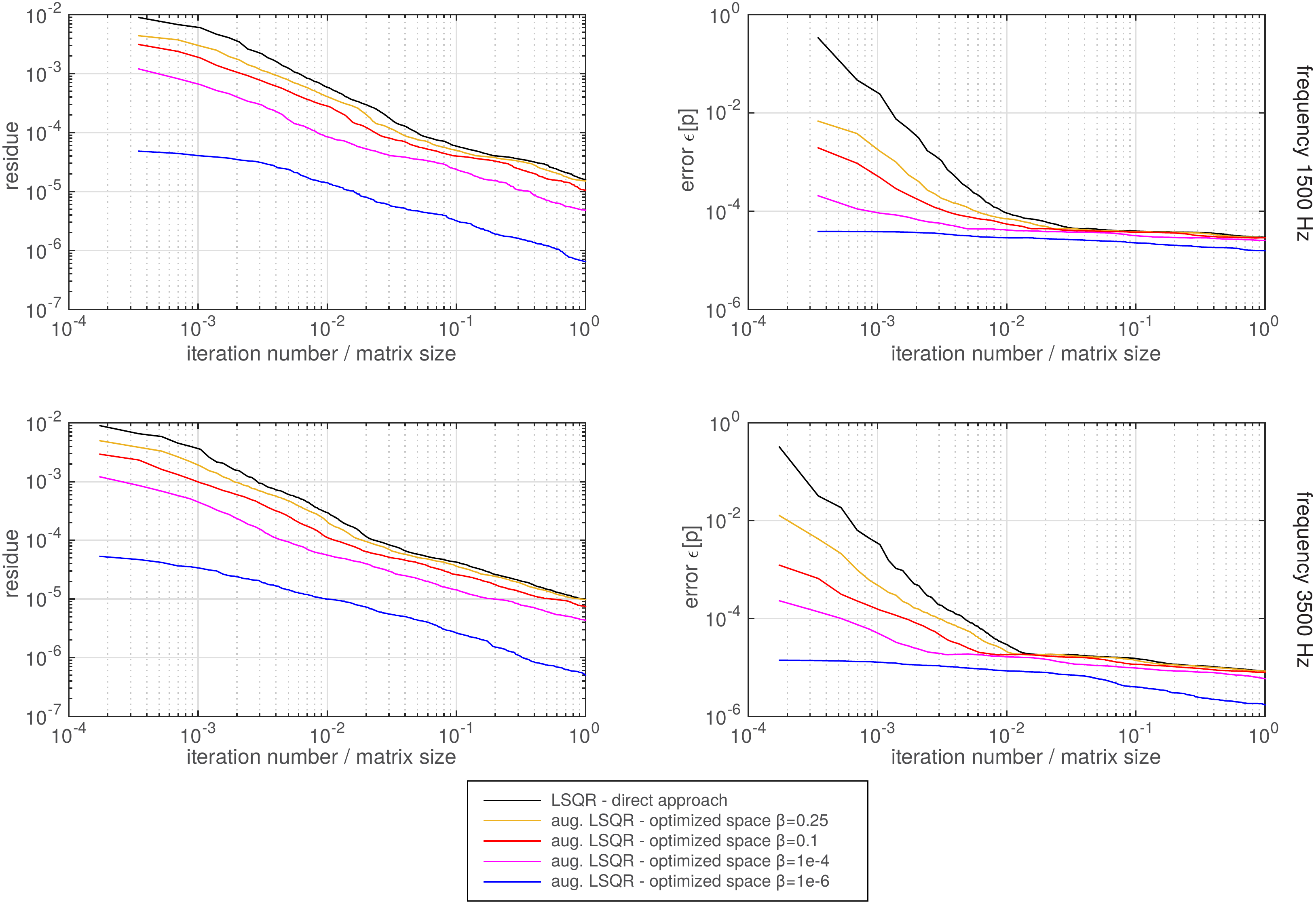} 
   \caption{Scattering by a sound-hard cylinder (Section \ref{subsection: scattering of a cylinder}): Evolution of the residu (left) and  the error $\varepsilon[p]$ (right) with the normalized number of iterations of the LSQR  at the f=1500Hz (top) and f=3500Hz (bottom) for the direct VTCR approach (black) or the O-VTCR with $\beta=0.25, 0.1, 10^{-4}$ and $10^{-6}$ (respectivelly orange, red, magenta and blue).}
   \label{iteration}
\end{figure}

We now study the convergence of augmented-LSQR solvers (for various $\beta$) compared to classical LSQR. The underlying discretization is sufficiently fine for a solution of good quality be obtained ($\mu=1.2$). Figure~\ref{iteration} shows the evolution of the norm of the residual and of the error in pressure during LSQR-iterations. 

We observe that in terms of residual, curves are almost linear and parallel: augmentation does not increase the convergence rate but  enables to start at a much smaller level of error. Note that in terms of pressure, the final precision of all methods practically coincides with the starting precision of A-LSQR with $\beta=1e^{-6}$.

\subsection{Sensitivity with respect to the mesh}

\begin{figure}[ht]
\centering
  \includegraphics[width=1\linewidth]{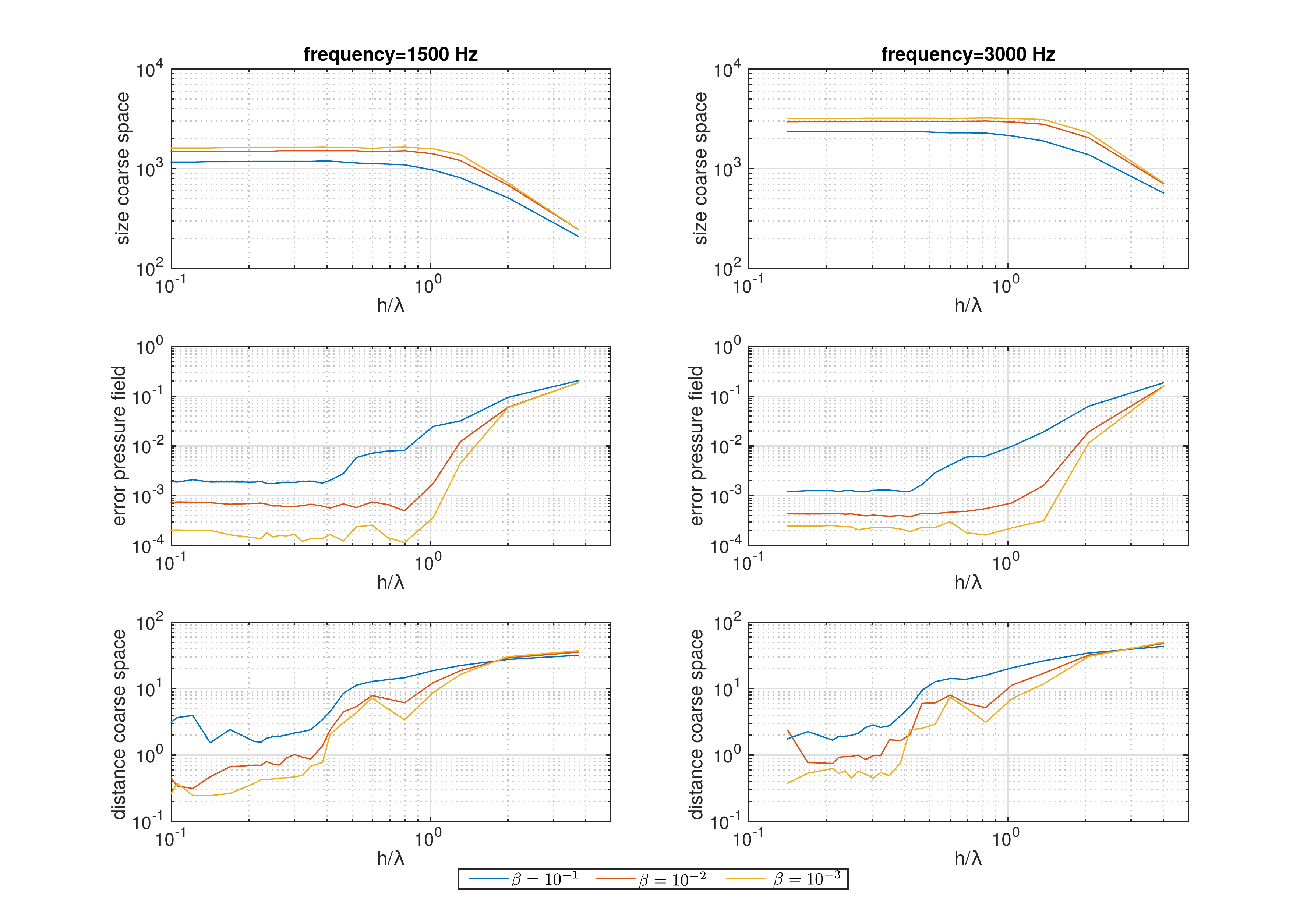} 
   \caption{Scattering by a sound-hard cylinder (Section \ref{subsection: scattering of a cylinder}): Dependence of the coarse space wrt the mesh (size, resulting precision, distance to reference coarse space) at frequencies 1\,500 Hz and 3\,500 Hz, and for various values of $\beta$.}
   \label{fig:mesheffect}
\end{figure}

In this subsection, we investigate the difficulties caused by the use of a mesh to interpolate the pressure and estimate its energy. The figure~\ref{fig:mesheffect} how the properties of the coarse subspace evolves according to the mesh refinement. The characteristic length of the mesh $h$ is normalized by the wavelength $\lambda$. Note that for $\frac{h}{\lambda}>\frac{1}{2}$, it is known that the interpolation error is not controlled which means that we analyze the left part of the plot ($\frac{h}{\lambda}\leqslant\frac{1}{2}$).

In the domain of interest, we observe (on the first row of Figure~\ref{fig:mesheffect}) that the dimension of the selected coarse space is almost constant (in particular for the finer $\beta$'s). More qualitatively, we measure the error in pressure from the resulting O-VTCR approach (second row of Figure~\ref{fig:mesheffect}). For $\beta\leqslant 10^{-2}$, the error reaches a plateau when the mesh is fine enough $\frac{h}{\lambda}\leqslant\frac{1}{3}$. The plateau is reached for finer meshes when $\beta=10^{-1}$. Of course, the value of the plateau depends on $\beta$.

Finally we propose a measure of the evolution of the subspace itself. Let $\tV_r$ be a (Euclidean-orthonormal) basis of the reference coarse space computed with the finest mesh, and $\tV_h$ be a  (Euclidean-orthonormal) basis of the coarse space computed with a mesh of dimension size $h$. Since dimension of the coarse space tends to increase with the refinement of the mesh, we propose the following measure for the distance between the spaces:
\begin{equation}
d = \|( \mathbf{I}-\tV_h\tV_h^H)\tV_r\|_{Fro}
\end{equation}
$d$ is thus the Frobenius norm of the projection of the reference space orthogonally to the current space. It somehow corresponds to the dimension of the supplementary subspace of $\operatorname{span}(\tV_h)$ in $\operatorname{span}(\tV_r)$. When normalized by the size of the coarse subspace, we see that as long as $h\leqslant \lambda$, the relative distance with the reference subspace is less than $1\%$.

This short study validates the choice $h=\frac{\lambda}{3}$ as a good instruction for the mesher, in particular when $\beta\leqslant 10^{-2}$.

\section{Numerical example: two dimensional car cavity} \label{subsection: car}

\begin{figure}[ht]
\centering
  \includegraphics[width=0.7\linewidth]{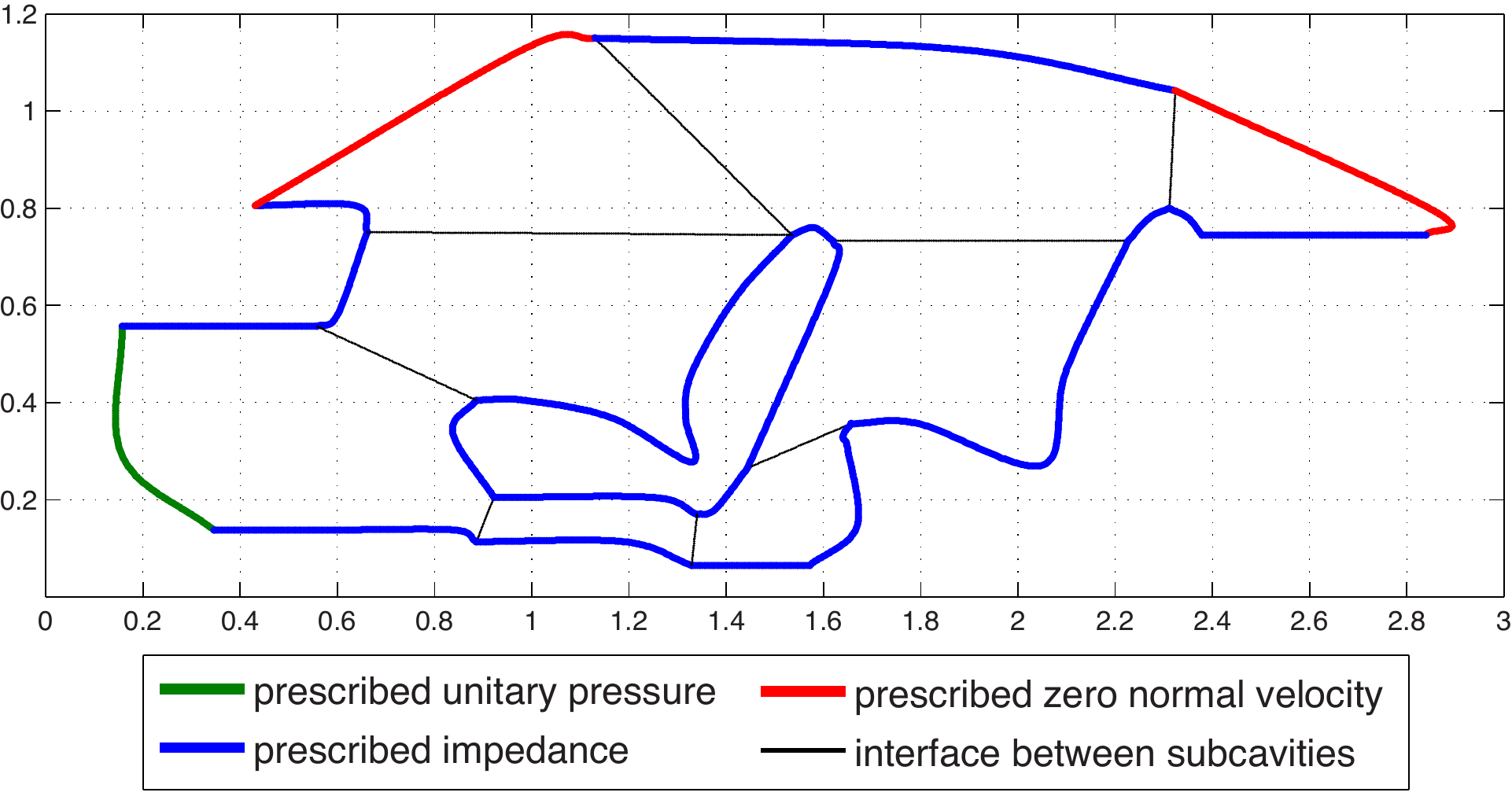} 
   \caption{Geometry and boundary conditions of the two dimensional car cavity.}
   \label{condilim}
\end{figure}

The VTCR, O-VTCR and A-VTCR are used to solve an acoustic problem for the car cavity depicted in Figure (Fig.~\ref{condilim}). 
The cavity is filled with air ($\rho$ = 1.25 kg.m$^{-3}$, $c$ = 330 m.s$^{-1}$, and $\eta=0$), it is excited by a uniform harmonic pressure in the front (boundary condition of type~(\ref{equation: reference problem}a) with $p_d=1$). The front and rear windows are hard walls (boundary condition of type~(\ref{equation: reference problem}c) with $v_d=0$), while an impedance condition of type~(\ref{equation: reference problem}b) is prescribed over all other boundaries (with $Z=\rmi (\rho \omega)/1245$).  The  cavity is decomposed into 8 sub-cavities.

\begin{figure}[ht]
\centering
  \includegraphics[width=1\linewidth]{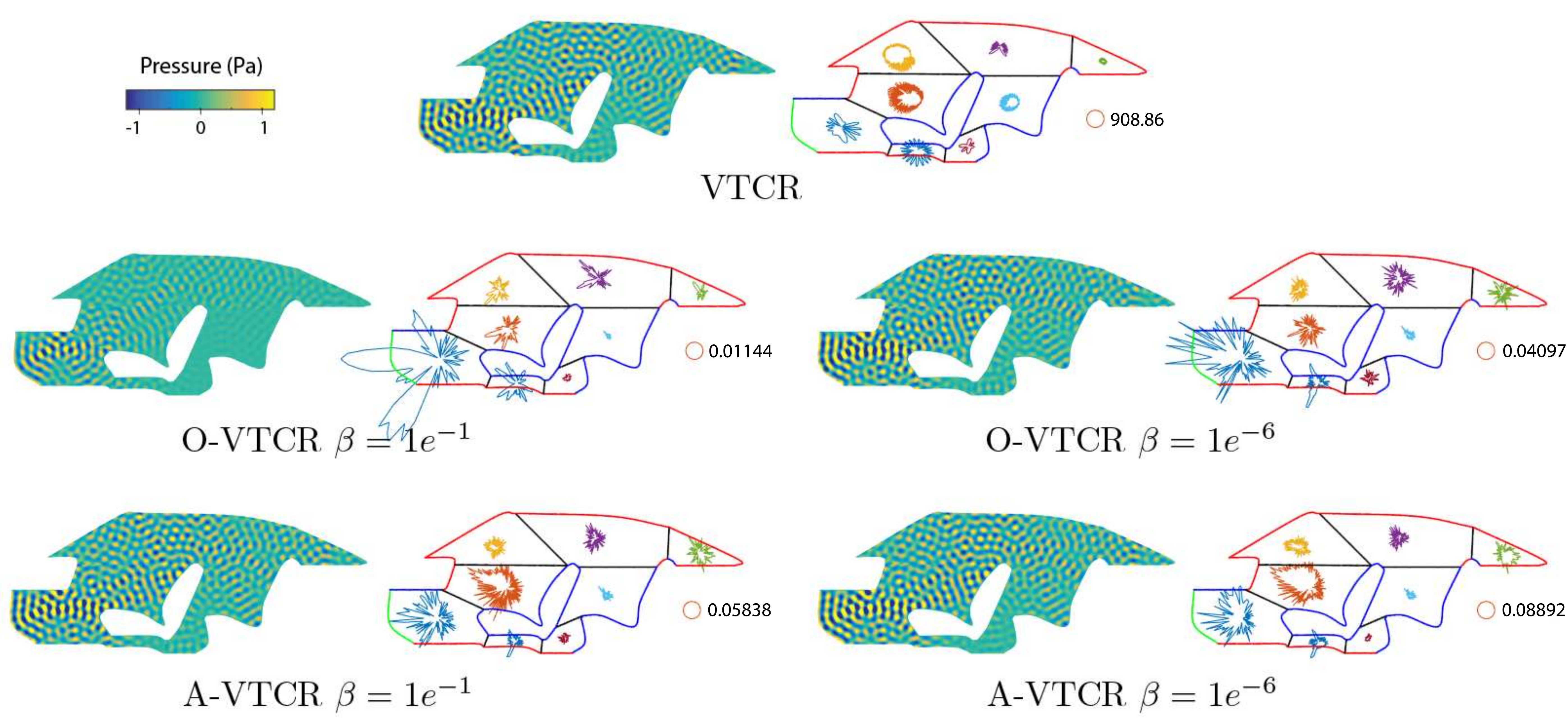} 
    \caption{Two dimensional car cavity (Section \ref{subsection: car}): pressure field and amplitude distribution at 5000Hz obtained with different resolution method: VTCR (top), O-VTCR (second line) and A-VTCR (third line), with $\beta=10^{-1}$ (left) and $\beta=10^{-6}$ (right) }
   \label{resultscar5000}
\end{figure}

\begin{figure}[ht]
\centering
  \includegraphics[width=1\linewidth]{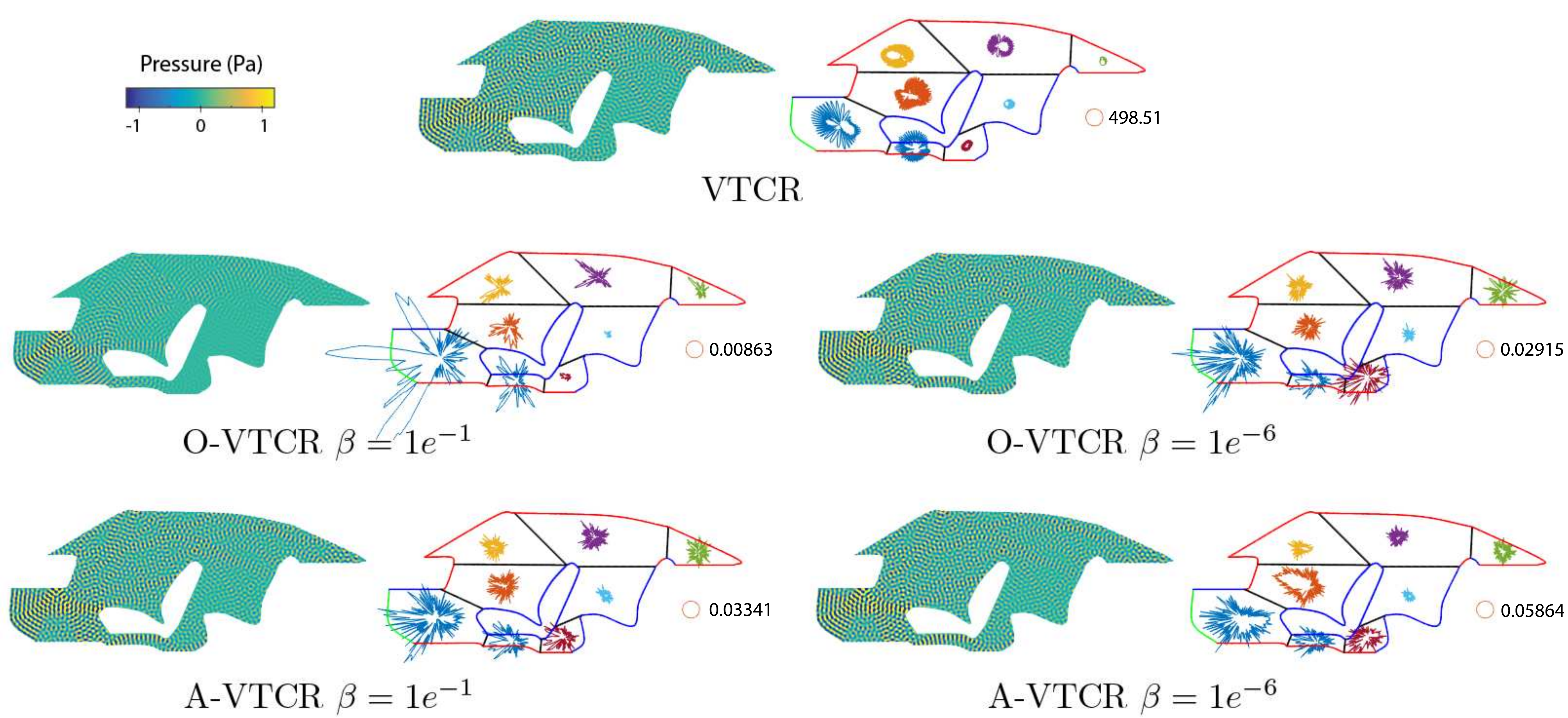} 
   \caption{Two dimensional car cavity (Section \ref{subsection: car}): pressure field and amplitude distribution at 10000Hz obtained with different resolution method: VTCR (top), O-VTCR (second line) and A-VTCR (third line), with $\beta=10^{-1}$ (left) and $\beta=10^{-6}$ (right) }
      \label{resultscar10000}
\end{figure}

\begin{figure}[ht]
\centering
  \includegraphics[width=1\linewidth]{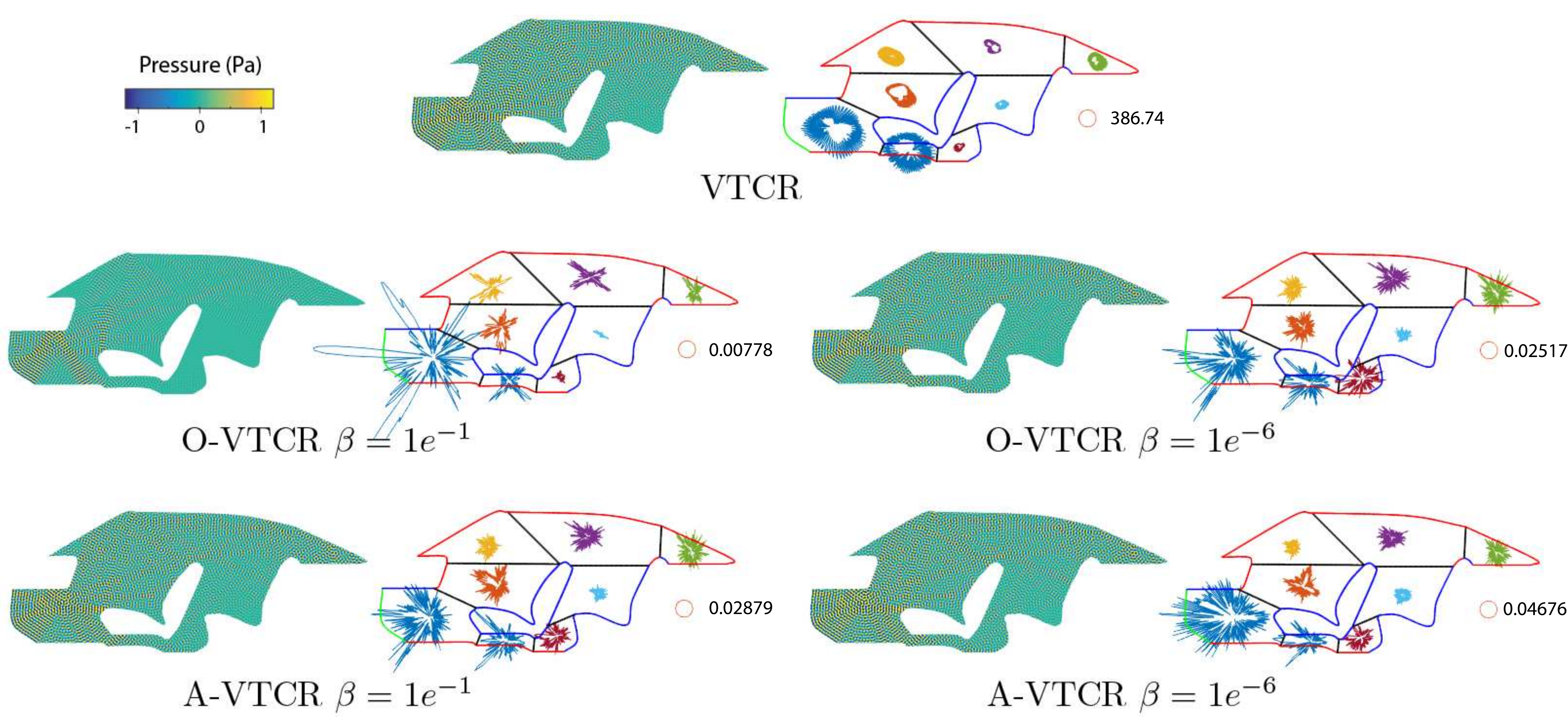} 
    \caption{Two dimensional car cavity (Section \ref{subsection: car}): pressure field and amplitude distribution at 15000Hz obtained with different resolution method: VTCR (top), O-VTCR (second line) and A-VTCR (third line), with $\beta=10^{-1}$ (left) and $\beta=10^{-6}$ (right) }
   \label{resultscar15000}
\end{figure}

The problem is solved at three different frequencies 3000 Hz, 10000 Hz and 15000Hz using either the classic VTCR with a pseudo inverse solver, the O-VTCR and A-VTCR with $\beta=10^{-1}$ and $10^{-6}$. The discretization is done using the approximation by Dirac's \eqref{eq:Dirac} associated with $\mu=1.5$, the integrals needed to compute the matrix $\mathbf{K}$ are performed numerically. The augmented LSQR has a stopping criterion of a relative residual of $10^{-3}$.
The calculated pressure field and amplitude distribution of each sub-cavity are represented in the figures \ref{resultscar5000},\ref{resultscar10000} and \ref{resultscar15000}. The scales used for the amplitude distribution are given by the circles on the right of the drawing corresponding to the different methods.

\begin{table} [ht]
\centering
\begin{tabular}{cc|c|c|c|}
\cline{3-5}
& &5000 Hz&10000 Hz  &15000 Hz  \\ 
\hline
\multicolumn{1}{|c|}{\multirow{2}{*}{VTCR}} & \multicolumn{1}{|c|}{size} & $1020 \times 1020$ & $ 2020 \times 2020$ &$ 3022 \times 3022 $\\
 \cline{2-5}
 \multicolumn{1}{|c|}{} & \multicolumn{1}{|c|}{condition number} & $8.9\, 10^{18}$&$ 3.47\, 10^{19}$&$ 5.6\, 10^{19}$\\ 
  \hline
  \hline
\multicolumn{1}{|c|}{\multirow{2}{*}{A/O-VTCR}} & \multicolumn{1}{|c|}{size} & $ 346 \times 346 $ & $ 684 \times 684 $ &$ 1022 \times 1022 $\\
 \cline{2-5}
 \multicolumn{1}{|c|}{} & \multicolumn{1}{|c|}{condition number} & $423$& $ 1.3\, 10^{3}$&$ 4.9\, 10^{3}$\\ \cline{2-5}
\multicolumn{1}{|c|}{  $\beta=10^{-1}$}  & \multicolumn{1}{|c|}{Relative error O-VTCR} & 0.305  &0.2934 &0.3311\\
 \cline{2-5}
 \multicolumn{1}{|c|}{} & Aug LSQR iterations & 98 & 103 & 143\\\cline{2-5}
\multicolumn{1}{|c|}{  }  & \multicolumn{1}{|c|}{Relative error A-VTCR} & 0.0036  &0.0629 & 0.0037  \\  
  \hline\hline
\multicolumn{1}{|c|}{\multirow{2}{*}{A/O-VTCR}} & \multicolumn{1}{|c|}{size} & $ 526 \times 526 $& $ 1008 \times 1008 $ &$ 1483 \times 1483 $\\
 \cline{2-5}
 \multicolumn{1}{|c|}{} & \multicolumn{1}{|c|}{condition number} & $ 1.8\, 10^{3}$&  $ 3.9\, 10^{3}$&$ 5.5\, 10^{3}$\\ \cline{2-5}
\multicolumn{1}{|c|}{  $\beta=10^{-6}$ }  & \multicolumn{1}{|c|}{Relative error O-VTCR} & 0.0116  & 0.0479& 0.0051\\ 
 \cline{2-5}
 \multicolumn{1}{|c|}{} & Aug LSQR iteration& 9 & 6& 18 \\ \cline{2-5}
\multicolumn{1}{|c|}{  }  & \multicolumn{1}{|c|}{Relative error A-VTCR} &0.0012  & 0.0438 & 0.0032  \\ 
  \hline
\end{tabular}
\caption{Two dimensional car cavity (Section \ref{subsection: car}): size and condition number of the algebraical system and relative error obtain with the different strategies }
\label{errorcar}
\end{table}

One can see that the pressure fields are very similar between the VTCR and the A/O-VTCR, with an exception for the O-VTCR with $\beta=0.1$. This visual impression is confirmed by the measure of the error given in the table \ref{errorcar} (the direct VTCR is considered to be the reference). However the distributions of amplitudes  are different. The ones obtained with the pseudo inverse have extremely large norms and therefore the identification of the main directions of propagation is impossible. Both O-VTCR and A-VTCR lead to much smaller amplitudes and relatively clear main directions of propagation. The iterations of A-VTCR lead to noisier portraits  of amplitudes than O-VTCR, especially at the highest frequency. 

The table \ref{errorcar} also presents the size and condition number of the algebraical system obtained with the different approaches. The condition number of the original systems is extremely high which corresponds to almost zero eigenvalues whereas the condition number of the reduced system is controlled around $10^3$. 
 As expected, for the O/A-VTCR, the precision is improved when reducing the value of $\beta$. One can observe that for the three considered frequencies, the A-VTCR leads to an accurate results even for large $\beta$.

\begin{table} [ht]
\centering
\begin{tabular}{cc|c|c|c|}
\cline{3-5}
& &5000 Hz&10000 Hz  &15000 Hz  \\ 
\hline
\multicolumn{1}{|c|}{\multirow{3}{*}{VTCR}} & \multicolumn{1}{|c|}{Matrix Assembly } & 0.34 s& 1.32 s& 3.82 s  \\ 
  \cline{2-5}
  \multicolumn{1}{|c|}{} &System solving& 0.95 s & 5.75 s& 15.32 s\\ \cline{2-5}
  \multicolumn{1}{|c|}{} &\bf Relative resolution time&\bf  1 &\bf  1& \bf 1\\
  \hline
  \hline
\multicolumn{1}{|c|}{\multirow{2}{*}{A/O-VTCR}} & \multicolumn{1}{|c|}{Construction Coarse space} & 0.14 s& 1.05 s& 2.02 s  \\   \cline{2-5}
  \multicolumn{1}{|c|}{} &Projection and Factorization& 0.25 s & 1.84 s &2.84 s \\
 \cline{2-5}
\multicolumn{1}{|c|}{  $\beta=10^{-1}$}  & \multicolumn{1}{|c|}{Total Resolution time O-VTCR} & 0.39 s& 2.89 s& 4.86 s  \\  \cline{2-5}
  \multicolumn{1}{|c|}{} &\bf Relative resolution time O-VTCR&\bf  0.41  & \bf 0.50& \bf 0.31 \\
     \cline{2-5}
  \multicolumn{1}{|c|}{} &Aug LSQR time& 0.98 s & 3.57 s & 5.49 s \\
   \cline{2-5}
\multicolumn{1}{|c|}{ }  & \multicolumn{1}{|c|}{Total Resolution time A-VTCR} & 1.37 s& 6.46 s& 10.37 s  \\  \cline{2-5}
\multicolumn{1}{|c|}{ }  & \multicolumn{1}{|c|}{ \bf Relative resolution time A-VTCR} &  \bf 1.44  & \bf 1.12 & \bf  0.67 \\ 
  \hline\hline
  \multicolumn{1}{|c|}{\multirow{2}{*}{A/O-VTCR}} & \multicolumn{1}{|c|}{Construction Coarse space} & 0.14 s& 1.05 s& 2.02 s  \\   \cline{2-5}
  \multicolumn{1}{|c|}{} &Projection and Factorization& 0.43 s & 2.74 s& 5.54 s\\
 \cline{2-5}
\multicolumn{1}{|c|}{  $\beta=10^{-6}$}  & \multicolumn{1}{|c|}{Total Resolution time O-VTCR} & 0.57 s& 3.79 s& 7.76 s   \\  \cline{2-5}
  \multicolumn{1}{|c|}{} & \bf Relative resolution time O-VTCR& \bf  0.6  & \bf 0.65&  \bf 0.51  \\
    \cline{2-5}
  \multicolumn{1}{|c|}{} &Aug LSQR time& 0.11 s & 0.45 s & 1.41 s \\
   \cline{2-5}
\multicolumn{1}{|c|}{ }  & \multicolumn{1}{|c|}{Total Resolution time A-VTCR} & 0.68 s& 4.24 s& 9.17 s \\  \cline{2-5}
\multicolumn{1}{|c|}{ }  & \multicolumn{1}{|c|}{ \bf Relative resolution time A-VTCR} & \bf  0.71  & \bf  0.73 & \bf  0.59 \\ 
\hline
\end{tabular}\caption{Two dimensional car cavity (Section~\ref{subsection: car}): CPU time required for the different steps of the resolution for the VTCR, O-VTCR and A-VTCR. }
\label{perfcar}
\end{table}

 The table \ref{perfcar} shows the computational time of the different steps of the resolution for the three methods at the three considered frequencies. For comparison, relative time with respect to classical direct VTCR is also provided. As one can see, the assembly of the VTCR matrix is relatively inexpensive in comparison to the resolution, making the resolution the bottleneck of the VTCR. 
 
 Note that our Matlab code is far from optimized. In particular the natural parallelism of the construction of the optimized space $(\At_{\sigma,h}\E)$ is not exploited, and the augmented LSQR solver is crudely implemented. The CPU time for O/A-VTCR could thus easily be  reduced. 
 
Yet we observe interesting performance for O-VTCR and A-VTCR. The speed-up of theses approaches relative to the direct VTCR seems to increase for large problems. This is due to the fact that direct solvers have a cubic complexity with respect to the size of the problem whereas the iterative solvers' complexity is quadratic (when the number of iterations remains relatively small). 

The sole cases where the new methods behave poorly, are when we try to achieve good precision with iterations starting with a poor coarse problem (A-VTCR with $\beta=10^{-1}$) on small problems ($f\leqslant 10000$ Hz).

In the end it seems that choosing a high precision coarse model is the most interesting strategy ($\beta=10^{-6}$). In that case, the coarse problem is two times smaller than the original system, the solution in the optimized subspace (O-VTCR) is already of good quality (for a computational time of $49\%$ of the reference time); if needed few iterations are sufficient to lower the residual (A-VTCR) at a limited cost (the gain in terms of time is then 41\%).

\section{Conclusion} \label{Conclusion}
When approximating Helmholtz equation, the Variational Theory of Complex Rays is a powerful alternative to the classical finite element because it is less subjected to the dispersion error. However the underlying representation of the unknown field in terms of amplitudes involves a compact operator which causes an accumulation of eigenvalues near zero, loss of coercivity and explosion of the condition number of the discrete system. 
In this paper we propose to filter the approximation space by creating a basis of amplitudes which have a significant contribution to the unknown in the appropriate norm. Note that the proposed method should adapt seamlessly to other methods which involve a compact operator in the representation of the unknown field. 

In practice, a mesh is introduced in order to handle the pressure fields generated by distributions of amplitudes; this mesh is much coarser than would be required for a reliable finite element computation, typically the mesh used to visualize the solution is sufficient. 

The selection process is associated with computing the highest part of the spectrum of a generalized eigenvalue problem independently on each subdomain. In most cases a simplification can be applied which leads to using singular value decomposition instead of generalized eigensystem. A truncation parameter must be introduced by the user, which controls the coercivity of the discrete sesquilinear form and the loss of precision with respect to the complete approximation space. This parameter is connected to the amount of energy represented in the filtered subspace. {Numerical experiments prove that the criterion makes the selection process almost independent of the mesh (assuming it is reasonably refined with respect to the wavelength).

The filtered subspace can be used to obtain a good quality approximation of the solution with a system of reduced size and good conditioning. It can also be used as the coarse grid of an augmented Krylov solver. In practice having $99.9999\%$ of the energy inside the coarse space seems a good choice since both conditioning and precision remain correct, if needed only few LSQR iterations will be required to achieve full precision. 

Assessments proved that the method gave interesting performance in terms of precision, stability of the amplitude solution and even CPU time (even with a non-optimized implementation). 
}

\bigskip
\textbf{Acknowledgement: }
The authors wish to thank Marc Bonnet and Martin Vorhalik for their helpful discussions.

\bibliographystyle{ieeetr}
\bibliography{tvrc}

\begin{thebibliography}{10}

\bibitem{zienkiewicz1977}
O.~C. Zienkiewicz, {\em The Finite Element Method}.
\newblock McGraw-Hill, 1977.

\bibitem{Deraemaeker1999}
A.~Deraemaeker, I.~Babuska, and P.~Bouillard, ``Dispersion and pollution of the
  {FEM} solution for the {H}elmholtz equation in one, two and three
  dimensions,'' {\em International Journal for Numerical Methods in
  Engineering}, vol.~46, pp.~471--499, 1999.

\bibitem{Moiola2014}
A.~Moiola and E.~A. Spence, ``Is the {H}elmholtz equation really
  sign-indefinite?,'' {\em SIAM Review}, vol.~56, no.~2, pp.~274--312, 2014.

\bibitem{dg12}
D.~A. Di~Pietro and A.~Ern, {\em Mathematical Aspects of Discontinuous
  {G}alerkin Methods}, vol.~69 of {\em Mathématiques et Applications}.
\newblock Springer, 2012.

\bibitem{gittelson2014dispersion}
C.~J. Gittelson and R.~Hiptmair, ``Dispersion analysis of plane wave
  discontinuous {G}alerkin methods,'' {\em International Journal for Numerical
  Methods in Engineering}, vol.~98, no.~5, pp.~313--323, 2014.

\bibitem{cessenatu}
O.~Cessenat and B.~Despres, ``Application of an ultra weak variational
  formulation of elliptic {PDE}s to the two-dimensional {H}elmholtz problem,''
  {\em SIAM Journal on Numerical Analysis}, vol.~35, pp.~255--299, 1998.

\bibitem{farhatu}
C.~Farhat, I.~Harari, and L.~Franca, ``The discontinuous enrichment method,''
  {\em Computer Methods in Applied Mechanics and Engineering}, vol.~190,
  pp.~6455--6479, 2001.

\bibitem{desmetu}
W.~Desmet, P.~Sas, and D.~Vandepitte, ``An indirect {T}refftz method for the
  steady-state dynamic analysis of coupled vibro-acoustic systems,'' {\em
  Computer Assisted Mechanics and Engineering Sciences}, vol.~8, pp.~271--288,
  2001.

\bibitem{ladevezeu}
P.~Ladev\`eze, ``A new computational approach for structure vibrations in the
  medium frequency range,,'' {\em Comptes Rendus Académie des Sciences Paris,
  s\'erie II}, pp.~849--856, 1996.

\bibitem{strouboulist}
T.~Strouboulis and R.~Hidajat, ``Partition of unity method for {H}elmholtz
  equation: q-convergence for plane-wave and wave-band local bases,'' {\em
  Applications of Mathematics}, vol.~51, pp.~181--204, 2006.

\bibitem{riou2008}
H.~Riou, P.~Ladevèze, and B.~Sourcis, ``The multiscale {VTCR} approach applied
  to acoustics problems,'' {\em Journal of Computational Acoustics}, vol.~16,
  no.~4, pp.~487--505, 2008.

\bibitem{kovalevsky12}
L.~Kovalevsky, P.~Ladev\`eze, H.~Riou, and M.~Bonnet, ``The variational theory
  of complex rays for three-dimensional {H}elmholtz problems,'' {\em Journal of
  Computational Acoustics}, vol.~20, 2012.

\bibitem{huttunen2008interaction}
T.~Huttunen, J.~Kaipio, and P.~Monk, ``An ultra-weak method for acoustic
  fluid–solid interaction,'' {\em Journal of Computational and Applied
  Mathematics}, vol.~213, no.~1, pp.~166--185, 2008.

\bibitem{huttunen2002computational}
T.~Huttunen, P.~Monk, and J.~P. Kaipio, ``Computational aspects of the
  ultra-weak variational formulation,'' {\em Journal of Computational Physics},
  vol.~182, no.~1, pp.~27--46, 2002.

\bibitem{Morse}
P.~M. Morse, {\em Theoretical Acoustics}.
\newblock Princeton university press, 1968.

\bibitem{dautraylions}
R.~Dautray and J.-L. Lions, {\em Mathematical Analysis and Numerical Methods
  for Science and Technology}.
\newblock Springer, 2000.

\bibitem{obb98}
J.~T. Oden, I.~Babuška, and C.~E. Baumann, ``A discontinuous hp finite element
  method for diffusion problems,'' {\em Journal of Compututaional Physics},
  vol.~146, no.~2, pp.~491--519, 1998.

\bibitem{colton01}
D.~Colton and R.~Kress, ``On the denseness of {H}erglotz wave functions and
  electromagnetic {H}erglotz pairs in {S}obolev spaces,'' {\em Mathematical
  Methods in the Applied Sciences}, vol.~24, pp.~1289--1303, 2001.

\bibitem{Herglotz04}
N.~Weck, ``Approximation by {H}erglotz wave functions,'' {\em Mathematical
  Methods in the Applied Sciences}, vol.~27, no.~2, pp.~155--162, 2004.

\bibitem{kovalevsky11}
L.~Kovalevsky, P.~Ladev\`eze, and H.~Riou, ``The {F}ourier version of the
  variational theory of complex rays for medium-frequency acoustics,'' {\em
  Computer Methods in Applied Mechanics and Engineering}, pp.~142--153, 2012.

\bibitem{desmetu2}
W.~Desmet, B.~van Hal, P.~Sas, and D.~Vandepitte, ``A computationally efficient
  prediction technique for the steady-state dynamic analysis of coupled
  vibro-acoustic systems,'' {\em Advances in Engineering Software}, vol.~33,
  pp.~527--540, 2002.

\bibitem{sourcis}
H.~Riou, P.~Ladevèze, B.~Sourcis, B.~Faverjon, and L.~Kovalevsky, ``An
  adaptive numerical strategy for the medium-frequency analysis of
  {H}elmholtz's problem,'' {\em Journal of Computational Acoustics}, vol.~20,
  2012.

\bibitem{doi:10.1137/070706616}
B.~Cockburn, J.~Gopalakrishnan, and R.~Lazarov, ``Unified hybridization of
  discontinuous {G}alerkin, mixed, and continuous {G}alerkin methods for second
  order elliptic problems,'' {\em SIAM Journal on Numerical Analysis}, vol.~47,
  no.~2, pp.~1319--1365, 2009.

\bibitem{Nguyen20093232}
N.~Nguyen, J.~Peraire, and B.~Cockburn, ``An implicit high-order hybridizable
  discontinuous {G}alerkin method for linear convection–diffusion
  equations,'' {\em Journal of Computational Physics}, vol.~228, no.~9,
  pp.~3232 -- 3254, 2009.

\bibitem{brenner}
S.~C. Brenner and R.~Scott, {\em The Mathematical Theory of Finite Element
  Methods}, vol.~15 of {\em Texts in Applied Mathematics}.
\newblock Springer, 2008.

\bibitem{Kamenski14}
L.~Kamenski, W.~Huang, and H.~Xu, ``Conditioning of finite element equations
  with arbitrary anisotropic meshes,'' {\em Mathematics of Computation},
  vol.~83, pp.~2187--2211, 2014.

\bibitem{lsqr82}
C.~C. Paige and M.~A. Saunders, ``{LSQR}: An algorithm for sparse linear
  equations and sparse least squares,'' {\em ACM Transactions on Mathematical
  Software}, vol.~8, no.~1, pp.~43--71, 1982.

\end{thebibliography}

\end{document}